\documentclass[preprint2]{aastex61}

\newcommand{\kms}{km s$^{-1}$}

\shorttitle{Star Clusters in NGC 4589}
\shortauthors{Lee, Jang \& Kang}

\usepackage{color}
\usepackage{multirow}
\usepackage{amsmath}
\usepackage{ulem}

\begin{document}

\title{ 
Star Clusters in the Elliptical Galaxy NGC 4589 \\
Hosting a Calcium-rich SN Ib (SN 2005cz)}

\author{Myung Gyoon Lee}
\affiliation{Astronomy Program, Department of Physics and Astronomy, Seoul National University, Gwanak-gu, Seoul 151-742, Korea}
\email{mglee@astro.snu.ac.kr}
\author{In Sung Jang} 
\affiliation{Leibniz-Institut  f{\"u}r Astrophysik Potsdam (AIP), An der Sternwarte 16, D-14482, Potsdam, Germany}
\author{Jisu Kang$^1$} 

\begin{abstract}
NGC 4589, a bright E2 merger-remnant galaxy,  
hosts the peculiar fast and faint calcium-rich Type Ib supernova (SN) SN 2005cz. 
The progenitor of Ca-rich SNe Ib has been controversial: it could be a) a young massive star with 6-12 M$_\sun$ in a binary system,
or b) an old low-mass star in a binary system that was kicked out from the galaxy center. 
Moreover, previous distance estimates for this galaxy have shown a large spread, ranging from 20 Mpc to 60 Mpc. 
Thus, using archival $Hubble$ $Space$ $Telescope$/ACS $F435W$, $F555W$, and 
$F814W$ images, we search for star clusters in NGC 4589 in order to help resolve these issues.
We find a small population of 
young star clusters with $25<V\leq27$ ($-7.1<M_V\leq-5.1$) mag and age $< 1$ Gyr in the central region at $R<0\farcm5$ ($<3.8$ kpc), 
thus supporting the massive-star progenitor scenario for SN 2005cz.  
In addition to young star clusters, we also find a large population of old globular clusters.
In contrast to previous results in the literature, we find that  the color distribution of the globular clusters is clearly bimodal.
The turnover (Vega) magnitude in the $V$-band luminosity functions of the blue (metal-poor) globular clusters is determined to be 
$V_0{(\rm max)}=24.40\pm0.10$ mag.
We derive the total number of globular clusters, $N_{\rm GC} =640\pm50$,
and the specific frequency,  $S_N =1.7\pm0.2$.
Adopting a calibration for the metal-poor globular clusters, 
$M_V({\rm max})=-7.66\pm0.14$ mag, 
we derive a distance to this galaxy: 
 $(m-M)_0=32.06\pm0.10({\rm ran})\pm0.15({\rm sys})$ 
($d=25.8\pm2.2$ Mpc).
\end{abstract}


\section{Introduction}

Type Ib supernovae (SNe Ib) are the remnants of the collapsed core of massive WC Wolf-Rayet stars 
which lost most of their outer hydrogen envelope (so they are sometimes called thin-stripped core collapse SNe) \citep{sma09}. 
Their spectra show a distinguishable He I 5876\AA~ line but little feature of silicon lines. 
Recently, a new type of SNe Ib called Ca-rich SNe Ib has been discovered in NGC 1032 \citep{per10}.
The spectra of the early phase of Ca-rich SNe Ib show He I lines, while their spectra of the late phase show Ca lines. 
Ca-rich SNe Ib are much fainter and show a faster decline rate in their light curves compared to  normal SNe Ib  (see the review by \citet{tau17}, in his section 5). 

Ca-rich SNe Ib host many interesting properties.
First, about 50\% of all Ca-rich SNe Ib are found in E or S0 galaxies \citep{per10,kas12,tau17}.  
It is difficult to reconcile this fact 
with the conventional concept of massive stars being the progenitors of normal SNe Ib.
Second, they are often found much further from the center of their host galaxies and they are found even in the intracluster or intragroup region. 
So they are sometimes called as homeless SNe \citep{kas12,fol15,lun17}. 
Third, \citet{lym16} found no evidence of globular clusters or dwarf galaxies at the position of known Ca-rich SNe in the $Hubble$ $Space$ $Telescope$ ($HST$) images, and they concluded that
the progenitors of these Ca-rich SNe must have come from somewhere else, 
offset from the current SN position.  
 
These results have made the origin of Ca-rich SN Ib progenitors controversial: whether it  is a) a young massive star with 6-12 M$_\sun$ in a binary system \citep{kaw10,gva17,mor17}, or b) an old low-mass star in a binary system that was kicked out from the galaxy center or elsewhere \citep{per11,fol15,lym16}.

One of the most popular targets to study the progenitors of Ca-rich SNe Ib has been
NGC 4589 hosting SN2005cz 
\citep{kaw10,per11,fol15}.
NGC 4589 is a bright X-ray-emitting elliptical 
galaxy with a LINER nucleus. 
It is the brightest member of a loose group of galaxies (Group No.107 in \citet{gel83}). 
The red integrated color
($(V-I)=1.18$) and the relatively faint F160W-band absolute magnitude of the surface brightness fluctuation (SBF) of NGC 4589 indicate that 
this galaxy is dominated by old stellar populations (\citet{jen03}, see their Figs. 2,3 and 4).
Basic properties of NGC 4589 are summarized in {\color{blue}\bf Table 1}. 

\begin{deluxetable*}{lcc}
\tabletypesize{\footnotesize}
\setlength{\tabcolsep}{0.05in}
\tablecaption{Basic Parameters of NGC 4589}
\tablewidth{0pt}
\tablehead{ \colhead{Parameter} & \colhead{Value}  & \colhead{Reference}}
\startdata
R.A.(2000) 	& $12^h37^m25.^s0$ &		 RC3	\\
Dec(2000) 	& $74\arcdeg11\arcmin31\arcsec$&  RC3	\\
Type		& E2	, LINER		& RC3	\\
Foreground extinction, $A_B$, $A_V$, and $A_I$ & 0.102, 0.077, and 0.042  & \citet{sch11} \\
Distance modulus, $(m-M)_0$ &  $32.06\pm0.18$  & GCLF, this study\\
Distance, d [Mpc] 	& $25.8\pm2.2$ & GCLF, this study\\
Image scale & 125 pc arcsec$^{-1}$  & This study \\
$B$ total magnitude, $B^T$	& $11.69\pm0.15$  & RC3 \\
$V$ total magnitude, $V^T$	& $10.73\pm0.15$  & RC3 \\
$B$ absolute magnitude, $M_B$	& $-20.47\pm0.23$  & RC3, This study \\
$V$ absolute magnitude, $M_V$	& $-21.41\pm0.23$  & RC3, This study \\
Position angle$^a$			& 89 deg (B), 92 deg (K)   & RC3 \\
$D_{25} (B)$				& $189\farcs70 \times 153\farcs66$  & RC3 \\
$D_{\rm total}(K)$		& $249\farcs00 \times 186\farcs75$  & \citet{jar03} \\
Effective radius, $R_{\rm eff}$	& $30\farcs4\pm3
\farcs0$ ($3800\pm400$ pc)  
& \citet{moe89} \\
Heliocentric velocity, $v_h$ & $1980\pm14$ \kms  &RC3 \\
\hline
\enddata
\tablenotetext{a}{ 
Position angles measured in the $B$ and $K$ band images, respectively.} 
\label{tab_basic}
\end{deluxetable*}

However, the distance to NGC 4589 is still uncertain. Previous distance estimates for this galaxy show a large spread (ranging from 20 Mpc to 60 Mpc), though distance estimates based on the SBF method show a much smaller spread \citep{dev84,fab89,wil97,ton01,jen03,the07,bla10}. 
Therefore, in this study, we estimate the distance to this galaxy using the luminosity functions of the globular clusters (GCLFs) detected in this galaxy \citep{har01,ric03,dic06,rej12}.
 
More interestingly, observations of NGC 4589 show  
several peculiar features. 
It hosts a dust disk that is aligned along the minor axis, and it is rotating fast around its major axis \citep{moe89}. 
Ionized gas emission is detected along the minor axis at $R< 20''$ from the galaxy center. \citet{moe89} found that both gas and stellar components show strong rotation with complex kinematics. 
They suggested that a gas-rich galaxy fell into NGC 4589 and formed a rotating dust disk in the inner region of the galaxy, 
and that NGC 4589 is in an advanced state of merging.  
A small amount of $H_2$ gas,
$9.1 \times 10^7 M_\sun$, 
was detected in this galaxy from CO observations \citep{sof93}. 
PAH emissions at $11.3\mu m$  
were detected in the position of the dust lanes,
and far-IR and $17\mu m$ PAH emissions were found in the more extended regions of NGC 4589 \citep{kan08,kan10}.
\citet{kan10} suggested that $11.3\mu m$ PAH features may be due to the gas brought in by an early merger, and the far-IR emission and $17\mu m$ PAH features are relics of a later merger.

If NGC 4589 had a recent wet merger, then there may be a population of young star clusters that can provide the massive-star progenitor for SN 2005cz. 
Motivated by this idea, we search 
for star clusters in 
NGC 4589 using $HST$ images in the archive,
and investigate their properties to tell whether any young to intermediate age clusters exist or not in this galaxy.  

To date, there is only one published paper on the star clusters in NGC 4589.
\citet{kun01} presented $VI$ photometry of  bright globular clusters with $21<V<24.5$ mag in NGC 4589 based on shallow $HST$/WFPC2 $F555W$ and $F814W$ images. They found no bimodality in the color distribution of these globular clusters, which is in contrast to the cases of other bright elliptical galaxies that show mostly strong bimodality (\citet{bro06,har17} and references therein).
This may be an intrinsic nature of NGC 4589 or due to the shallow photometry in their study.
Thus, nothing is known about any of the young star clusters in this galaxy.

This paper is organized as follows.
Section 2 describes how we select star clusters from the $HST$ images of NGC 4589.
In \S3, we present photometric properties of the detected star clusters as well as their spatial and radial distributions. We derive their GCLFs, 
and use them to determine the distance to NGC 4589.
In \S4, we discuss the origin of SN2005cz in relation with our results, compare our GCLFs with the previous study, and compare our distance estimate with previous distance estimates.
Finally we summarize the main results in the conclusion.


\begin{figure*} 
\centering
\includegraphics[scale=1.5]{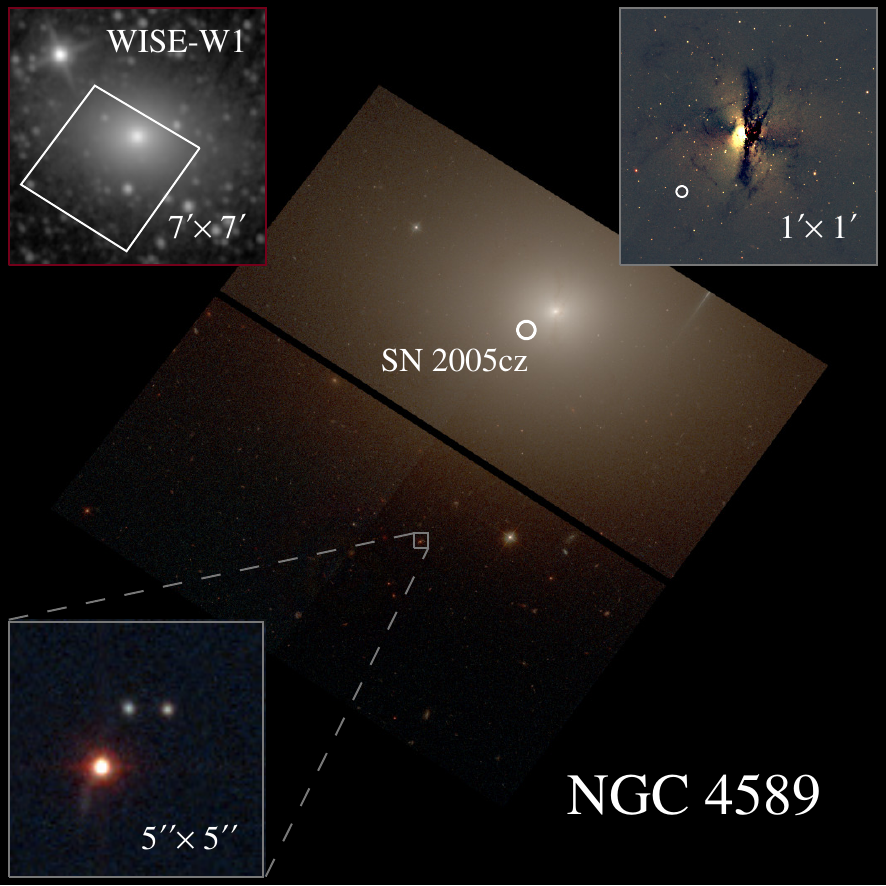} 
\caption{$HST$ color images of NGC 4589 hosting SN 2005cz (circles). North is up, and East is to the left. (Upper left) $WISE$ $W1$ image for $7'\times7'$ field.
(Upper right) A zoomed-in $1'\times1'$ $HST$ image of the central region from which galaxy light model was subtracted.
(Lower left) A zoom-in $5''\times5''$ image of a halo field shows two globular clusters candidates (two white sources). 
}
\label{fig_finder}
\end{figure*}

\section{Data Reduction and Star Cluster Selection}

\subsection{Data}

We used ACS images for NGC 4589 from the 
$HST$ archive (PI:Smartt, ID : 10498). The exposure times are 
1500s for $F435W$, 1500s for $F555W$, and 1600s for $F814W$. 
We combined individual images of NGC 4589 within the same filter using AstroDrizzle \citep{gon12}.
The image scale of the combined images is $0\farcs05$ per pixel.

{\color{blue}\bf Figure \ref{fig_finder}} displays 
a color image of the $HST$ field, the location of which is marked in  $WISE$-$W1$ (3.6 $\mu$m) image in the upper left. 
SN 2005cz is located at $\sim13''$ (1.6 kpc for the adopted distance of 25.8 Mpc here)
South-East from the galaxy center, as marked by the circle close to the galaxy center.
In the upper right, we display a zoomed-in image of the central region of NGC 4589 
from which galaxy light model was subtracted. Note the presence of dust lanes that are perpendicular to the major axis (that is almost horizontal in the image) in the central region. 
A zoomed-in image of a southern field in the lower left shows two compact sources, which are star cluster candidates in NGC 4589 (the red bright source below is a foreground star). 

\subsection{Data Reduction}
We  reduced the data following the procedures used in the study of star clusters in Coma galaxies by \citet{lee16}. 
We derived model images of the galaxy light using IRAF/ellipse after masking all the bright sources except 
NGC 4589.
Then we subtracted the model galaxy images from the drizzled images for better source detection. The resulting images are used for source detection and photometry with DAOPHOT \citep{ste94}. 
A master source list was made from the $F814W$ image with a detection threshold of $5\sigma$. 

Effective radii of typical globular clusters are 2--3 pc,
so globular clusters at the distance of NGC 4589 are expected to appear as point sources or as slightly resolved sources in the $HST$/ACS images 
with an image scale of $\sim$6 pc per pixel.
The magnitude difference between small and large apertures is one of the most effective central concentration parameters to distinguish between point sources and extended sources \citep{whi99,pen11}, as shown in the study of  star clusters using the $HST$ images for Coma galaxies in \citet{lee16}.
We calculated the values of the 
$F814W$ magnitude concentration parameter for the detected sources using aperture magnitudes with radii of 1.5 pixels and 3.0 pixels, $C{\rm (1.5 pix -3.0 pix)}$.

We derived $F814W$ aperture correction values 
for the 
sources including star cluster candidates following the method as used in \citet{lee16}.
The steps are as follows. First, we selected isolated bright sources with a various range of concentration parameter.
For these selected sources, we calculated the value of the magnitude difference between the aperture radii of 10 and 4 pixels,
$\Delta({\rm 10 pix -4 pix})$.
From linear fitting, we obtain
$\Delta({\rm 10 pix -4 pix}) = -0.564 C{\rm (1.5 pix -3.0 pix)} +0.175$ with $rms=0.028$.
We applied this aperture correction to derive a magnitude for the aperture radius of 10 pixels ($=0\farcs5$).
Then we derived $F435W$ and $F555W$ magnitudes
from $F814W$ magnitudes
using the 4 pixel radius colors of the sources, e.g., 
$F555W$ (10 pix) = $F814W$ (4 pix) + $\Delta(\rm 10 pix -4 pix)$ + ($F555W$ -- $F814W$)(4 pix).
Finally, we applied a further aperture correction for radii
$=0\farcs5$ to infinity using the values provided by the STScI (--0.106 mag for $F435W$, 
--0.096 mag for $F555W$ and --0.098 mag for $F814W$).
 
If we use aperture corrections for radii = 4  to 10 pixels derived for each band and apply them to obtain their colors, the resulting colors will have larger errors. The method adopted in this study is valid only 
if there is no color-gradient in the 4--10 pixel region of the sources.

According to the enclosed energy distribution of point sources in ACS data listed in Table 3 of \citet{sir05}, 
the stellar fluxes within a certain radius have a slight variation depending on the filter.
The fractions of stellar flux within 4 pixel radii aperture in this table
are 0.832,    0.843,  and   0.833 for $F435W$, $F555W$, and $F814W$, respectively.
These flux variations lead to very small color differences: 
$\Delta(F435W-F555W)= +0.014$ mag (=$-2.5\times log_{10}(0.832/0.843)$),  
and 
$\Delta(F555W-F814W)= -0.013$ mag (=$-2.5\times log_{10}(0.843/0.833)$).
We ignored this small color correction before deriving the total magnitudes of the $F555W$ and F$814W$ bands.

The instrumental magnitudes in the $HST$ system were converted to the standard calibrated $BVI$ magnitudes in the Johnson--Cousins system using the information in \citet{sir05}. The photometric zero-points ($c0$) and color terms ($c1$) we used are: 
$c0 = 25.842$ and $c1 = -0.089$ for $F435W$, 
$c0 = 25.704$ and $c1 = -0.054$ for $F555W$, and
$c0 = 25.495$ and $c1 = -0.002$ for $F814W$.
Transformation uncertainties are estimated to be $\pm0.02$ mag in each band. 

However, we noted that the photometric zero-points for ACS/WFC have a slight dependence on time. 
The photometric transformation in \citet{sir05} was made using $HST$ data for NGC 2419 and 47 Tuc taken in 2002, but observations for NGC 4589 were obtained in 2006. According to the STScI webpage\footnote{http://www.stsci.edu/hst/acs/analysis/} the photometric zero-points for $F435W$, $F555W$, and $F814W$ in 2006 are on average 0.03 mag smaller than those of in 2002. 
This variation in zero-points could lead to larger systematic uncertainties in the transformation. We, therefore, adopt a conservative value of $\pm0.03$ mag for the final uncertainty associated with the photometric transformation.
In this study we use Vega magnitudes
and use the `0' subscripts for extinction-corrected quantities in the following analysis.

\subsection{Size Estimation}
We estimated the effective radii of bright 
sources with $V\le25$ using the ISHAPE program \citep{lar99}.
A point spread function (PSF) modeling is one of the most important steps 
in the ISHAPE run. In $HST$ data, synthetic PSFs (e.g. TinyTim PSFs \citep{kri11}) are optimized for individual frame images. Individual frame images (*\_flc.fits) have a strong geometric distortion, so using synthetic PSFs  may not be the best choice (see ISHAPE manual for details).
For this reason many previous studies used drizzled images, which are geometric distortion corrected and mostly co-added images, with empirical PSFs. However, modeling empirical PSFs is not always easy. Most  extragalactic $HST$ fields have a limited number of  isolated   bright stars and they have a stellar spectral energy distribution (SED) different  from target star clusters. Selecting a clean point source is also not easy because of the presence of blended stars, compact star clusters, and compact background galaxies. All these difficulties are possible sources of uncertainties in the size estimation.

Thus we adopted an alternative approach solving most of the problems mentioned above by using TinyTim PSFs with a single-drizzled image.
The steps are as follows.
We generated 400 TinyTim PSFs 
and placed them onto chip 1 (science extension 4) and chip 2 (science extension 1) of an flc image. We set a spectral type of K4V for the PSFs, similar to those of old globular clusters.
We then drizzled this single frame flc image. The output image is corrected for the geometric distortion, but is not co-added.
This image was used for generating the input PSFs using IRAF/DAOPHOT for the ISHAPE run.
The drizzled images we used for the aperture photometry are co-added, so that they are not ideal for size estimation 
 based on the modelled PSFs.

We prepared F814W-band single drizzled images (from one flc image) of the original NGC 4589 data and used them for size estimation. We used a King model with a concentration parameter of 30 
to fit sources. 
The mean size for each stellar object was computed from individual frame measurements with a median-based $\sigma$-clip algorithm 
set at 2$\sigma$.
We assigned the standard deviation of the clipped size values for size estimation errors.
Angular radii were converted to linear radii, adopting a distance of
25.8 Mpc as 
derived in the following section.

\begin{figure} 
\centering
\includegraphics[scale=0.9]{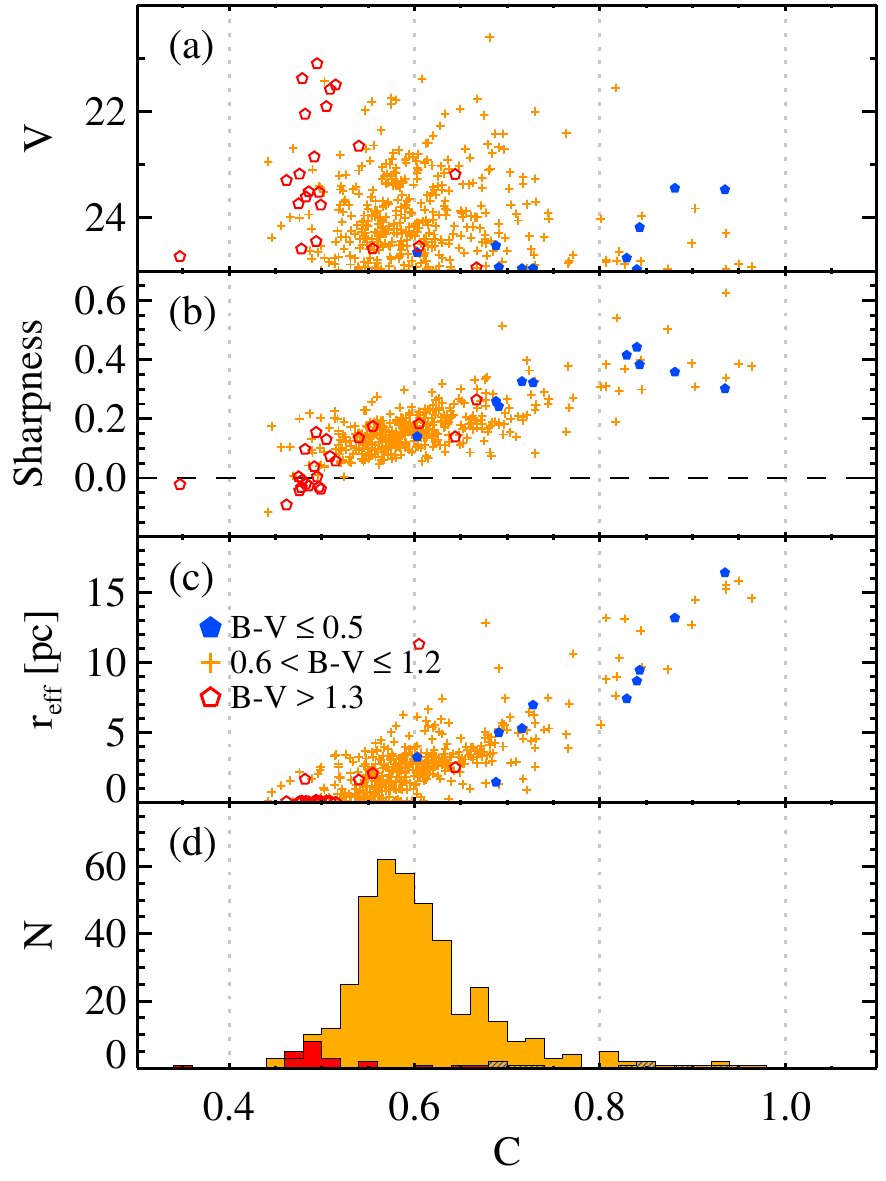}
\caption{Characteristics of the magnitude concentration parameter $C{\rm (1.5 pix -3.0 pix)}$ for the clean sample of the sources with $V\leq25$ mag selected through visual inspection (452 sources out of 546 sources):  
(a) $V$ magnitudes, 
(b) DAOPHOT sharpness parameters, 
(c) ISHAPE effective radii, and  
(d) the number distribution of $C$ for the bright sources. 
Blue, orange and red symbols represent the blue ($(B-V)\le0.5$), yellow ($0.6<(B-V)\le1.2$), and red (($B-V)>1.3$) sources, respectively.
The bright red sources show a strong concentration at $C\approx 0.5$, indicating that they are mostly stars. Note that the entire sample is dominated by the orange sources, and they show a strong peak at $C\approx 0.6$. They are mostly star clusters in NGC 4589. 
The blue sources are much fainter than the others, and they have, on average, a larger value of $C$ than the other groups.
They are mostly background galaxies.
}
\label{fig_C}
\end{figure}

\subsection{Star Cluster Selection}

Before selecting star cluster candidates in the list of the detected sources, we visually  inspected the images of all detected sources with $V(\rm total) \le27$ mag, 
and removed artifacts, blended sources, and sources with irregular morphology.
We set this magnitude limit 
by considering the photometric depth needed to cover a much fainter magnitude than the peak of the GCLF and the capacity of our visual inspection. Visual inspection of the images becomes difficult for the fainter sources. We selected 745 out of 2973 inspected sources.
This is the star cluster search sample. 

We prepared another sample that includes only bright sources with $V\le25$ mag.  We selected 452 out of 546 inspected sources with $V \le 25$ mag. 
The sources in the bright sample have smaller photometric errors and  are better for analysis than the fainter sources. We use this bright sample for the following analysis
of structural parameters and colors of the sources.

\begin{deluxetable*}{lcccccccccccc}
\tabletypesize{\scriptsize}
\setlength{\tabcolsep}{0.05in}
\tablecaption{A $BVI$ photometric Calatog of the Star Clusters  with $V\le 27$ mag in NGC 4589}
\tablewidth{0pt}
\tablehead{ \colhead{ID} & \colhead{R.A.}  & \colhead{Dec.} & \colhead{$V$} & \colhead{err($V$)} & \colhead{$(B-V)$} & \colhead{err($B-V$)} & \colhead{$(V-I)$} & \colhead{err($V-I$)} & \colhead{$r_{\rm eff}$} & \colhead{err($r_{\rm eff}$)} & \colhead{$C_I^a$} & \colhead{Remarks$^b$}\\
& (J2000) & (J2000) & [mag] & & & & & & [pc] & [pc]}
\startdata
    1    &189.5305461     &74.1714454    &26.002      &0.194      &1.100      &0.350      &0.802      &0.162     &...    &...     &0.705   &  \\
    2    &189.5239009     &74.1689122    &26.014      &0.134      &0.560      &0.191      &1.332      &0.119     &...    &...     &0.691   &  \\
    3    &189.5192765     &74.1694785    &23.733      &0.018      &1.394      &0.046      &2.227      &0.017       &0.02      &0.05     &0.475   &  \\
    4    &189.5187518     &74.1788576    &24.658      &0.050      &0.748      &0.069      &0.942      &0.041       &0.65      &1.05     &0.603   &  GC \\
    5    &189.5139799     &74.1710371    &25.916      &0.198      &0.067      &0.141      &0.256      &0.143     &...    &...     &0.705   &  \\
\hline
\enddata
\tablenotetext{a}{Concentration parameters derived from the F814W image.}
\tablenotetext{b}{Globular cluster candidates are marked by 'GC'.}
\tablecomments{Table \ref{tab_catalog} is published in its entirety in the electronic edition.  
The five sample star clusters  are shown here regarding its form and content.}
\label{tab_catalog}
\end{deluxetable*}

{\color{blue}\bf Figure \ref{fig_C}(a)} 
displays 
$V$-band magnitudes versus $C$ for the selected bright sources with $V\leq25$ mag. 
We divided the detected sources according to their color: the blue sources with $(B-V)\le0.5$, the globular cluster-like sources with $0.6<(B-V)\le1.2$, and the red sources with $(B-V)>1.3$. 
We adopted the color intervals for globular cluster selection based on the color distributions of  Milky Way globular clusters \citep{har96}. 
We chose our color selection intervals for the blue and red sources to have the cleanest and least contaminated sample of blue sources and red sources. 
{\color{blue}\bf Figures \ref{fig_C}(b) and (c)} 
display, respectively, 
the DAOPHOT sharpness parameter values and effective radii versus $C$.
In {\color{blue}\bf Figure \ref{fig_C}(d)} we plotted the $C$ number distributions for these sources.

Several features are noted in {\color{blue}\bf Figure \ref{fig_C}}. 
First, there is a strong concentration of sources at $C \approx 0.5-0.7$. They are dominated by sources with globular cluster-like colors.
Most of these sources are slightly resolved sources.
Thus these sources are mostly globular clusters in NGC 4589.
The median value of the effective radii of the globular clusters in NGC 4589 in this study is 2.5 pc (rms = 1.7 pc), which is similar to the value for the Milky Way globular clusters, 3 pc \citep{har96}. 
Second, the red sources show a narrow vertical plume at $C \approx 0.5$ in {\color{blue}\bf Figure \ref{fig_C}(a)}. The sources with  $C \approx 0.5$ have 
a median DAOPHOT sharpness value of 0.004 ($rms$ = 0.072), which shows that they are point sources. 
These point sources are dominated by red dwarf stars in the Milky Way.
Third, the number of blue sources is small. Blue sources show a broad distribution of $C$, and most of them have, on average, much larger $C$ values and fainter magnitudes than the other two groups.  They are considered to be mostly background galaxies. 
Fourth, the values of sharpness and effective radii show a strong correlation with the values of $C$. This shows that $C$ is a very effective parameter to distinguish point sources and extended sources.

We select, as the initial star cluster candidates, the slightly extended  sources with $0.52<C\le 0.8$ (referred to as the compact sources hereafter)
and the point sources with $C\le 0.52$. 
The $C$ distribution of the red sources, which are mostly foreground stars, shows a peak at $C\approx 0.5$, and declines to a zero value at $C>0.52$. On the other hand, the $C$ distribution of the globular cluster-like sources shows a peak at $C\approx 0.6$, and declines to a zero value at $C>0.8$. Therefore we chose $0.52<C\le0.8$ to have the cleanest and least contaminated GC sample. 
The selected cluster candidates are composed of mainly slightly resolved star clusters and a small number of point sources.
The selected point sources are considered to be mostly unresolved star clusters or stars.

In {\color{blue}\bf Table  2} 
we present a catalog of the star clusters in NGC 4589, including their $BVI$ photometry and effective radii.

\begin{figure*} 
\centering
\includegraphics[scale=0.9]{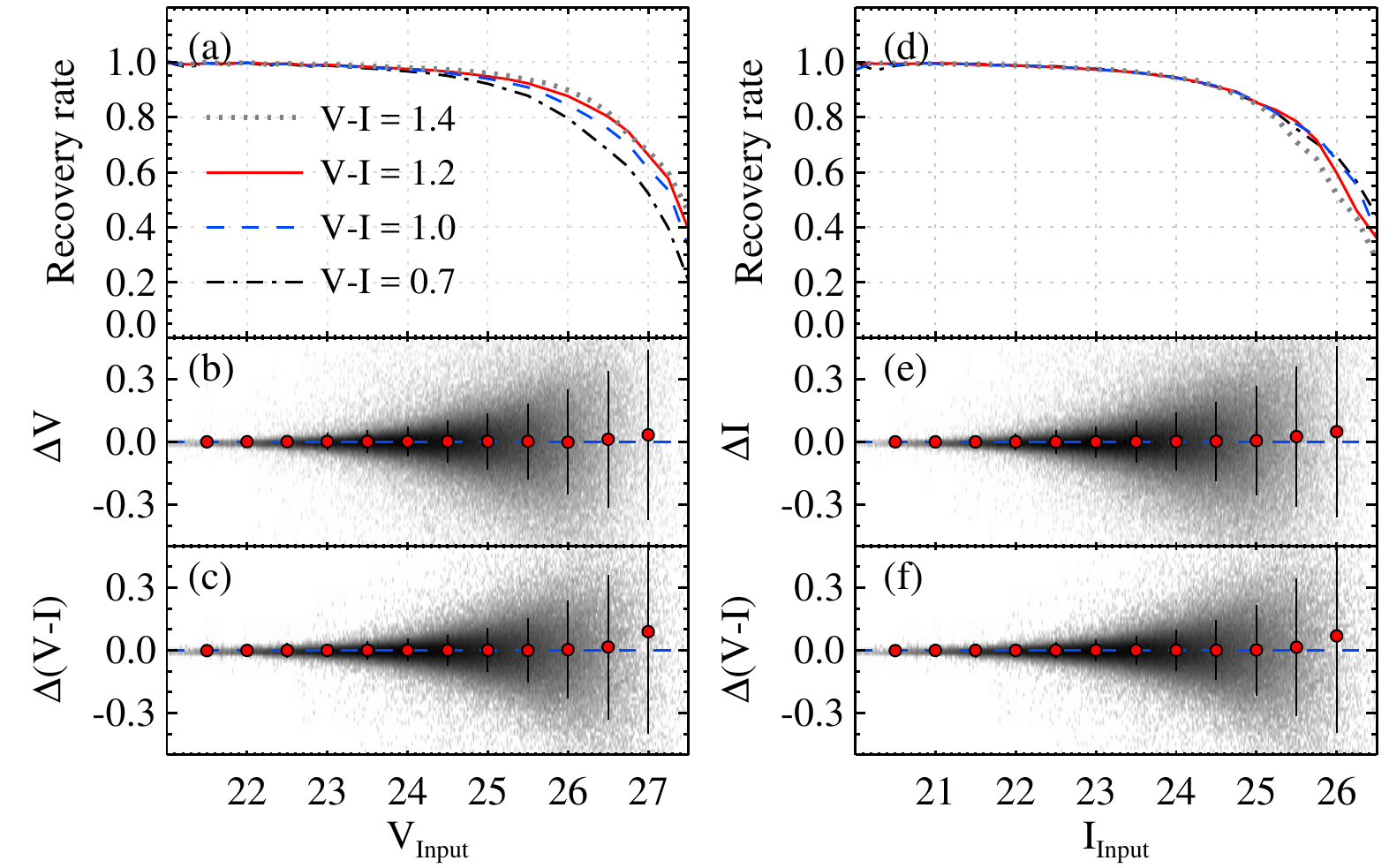}
\caption{Results of the completeness tests for $V$ (left panels) and $I$ (right panels) bands. 
Recovery rates (top panels), input magnitudes minus output magnitudes (middle panels), and input colors minus output colors (bottom panels) vs. magnitudes. 
Lines in the top panels
are for $(V-I)=0.7$, 1.0, 1.2, and 1.4 
($(B-V)=0.2, 0.75$, 0.95, and 1.4), respectively.
Red circles with error bars in the middle and  bottom panels denote the mean values with $\pm1 \sigma$ for given magnitudes.
The input sources have a magnitude range of $21 <V < 28$ mag (corresponding to $20 <I < 27$ mag for $(V-I)=1.0$).
}
\label{fig_comp}
\end{figure*}

\subsection{Completeness Tests}
We estimated the completeness of our photometry using artificial sources.
We generated images of the star cluster-like sources with $0.55<C\leq0.65$. 
We assumed that 
the luminosity function of the sources
is Gaussian with a peak at $V-$band total magnitude of 24.5 mag 
and a width of 1.0 mag, 
similar to the luminosity function of the globular clusters in NGC 4589 
derived in the following section. 
We adopted 
four colors 
for the artificial sources: 
$(B-V)=0.2$ and 1.4 $((V-I)=0.7$ and $1.4)$ which are close to the mean colors of the young and very red star clusters, and 
$(B-V)=0.75$ and 0.95 $((V-I)=1.0$ and $1.2)$ 
which are close to the mean colors of the blue and red globular clusters. 

We injected 250 artificial clusters onto the original image to create a test image. We repeated this procedure 1000 times, and we prepare 1000 test images.
We set the artificial sources to have a centrally concentrated spatial distribution.
The central region at $R<0\farcm1$ is masked out. 
We analysed these test images using the same procedures as used for the original images
in order to estimate the recovery rates,
 the ratios of  the number of recovered sources with respect to the number of input sources.

{\color{blue}\bf Figure \ref{fig_comp}} displays the results of the artificial source experiments.
The recovery rates for $V=27$ mag are 52\%, 61\%, 66\% and 68\% for (V-I) = 0.7, 1.0, 1.2, and 1.4, respectively. 
The mean values of the input magnitudes minus the output magnitudes are 
smaller than 0.03 mag for $V\leq26.5$ mag and $I\leq25.5$ mag. 

\begin{figure} 
\centering
\includegraphics[scale=0.9]{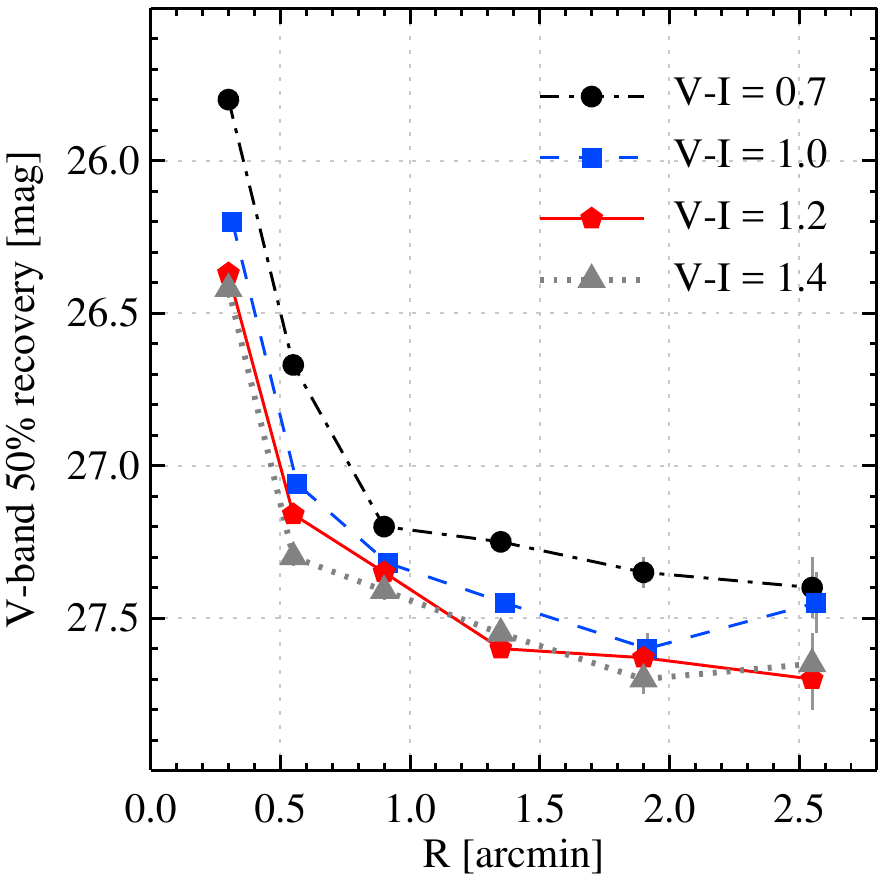}
\caption{
The 50\% completeness level as a function of the galactocentric distance. 
Circles, squares, pentagons, and triangles 
are for $(V-I)=$ 0.7, 1.0, 1.2, and 1.4, respectively. The central region at $R<0\farcm1$ was masked out.
}
\label{fig_comp2}
\end{figure}

In {\color{blue}\bf Figure \ref{fig_comp2}} we plot the $V$-band 50\% completeness level as a function of galactocentric distance.
The 50\% recovery magnitudes become fainter as  galactocentric distance increases. For $(V-I)=1.0$, the 50\% recovery magnitude is $V=26.2$ mag at $R=0\farcm3$, $V=27.3$ mag at $R=0\farcm9$, and $V\approx 27.5$ at $R=1\farcm3-2\farcm6$. 

\begin{figure*} 
\centering
\includegraphics[scale=1.0]{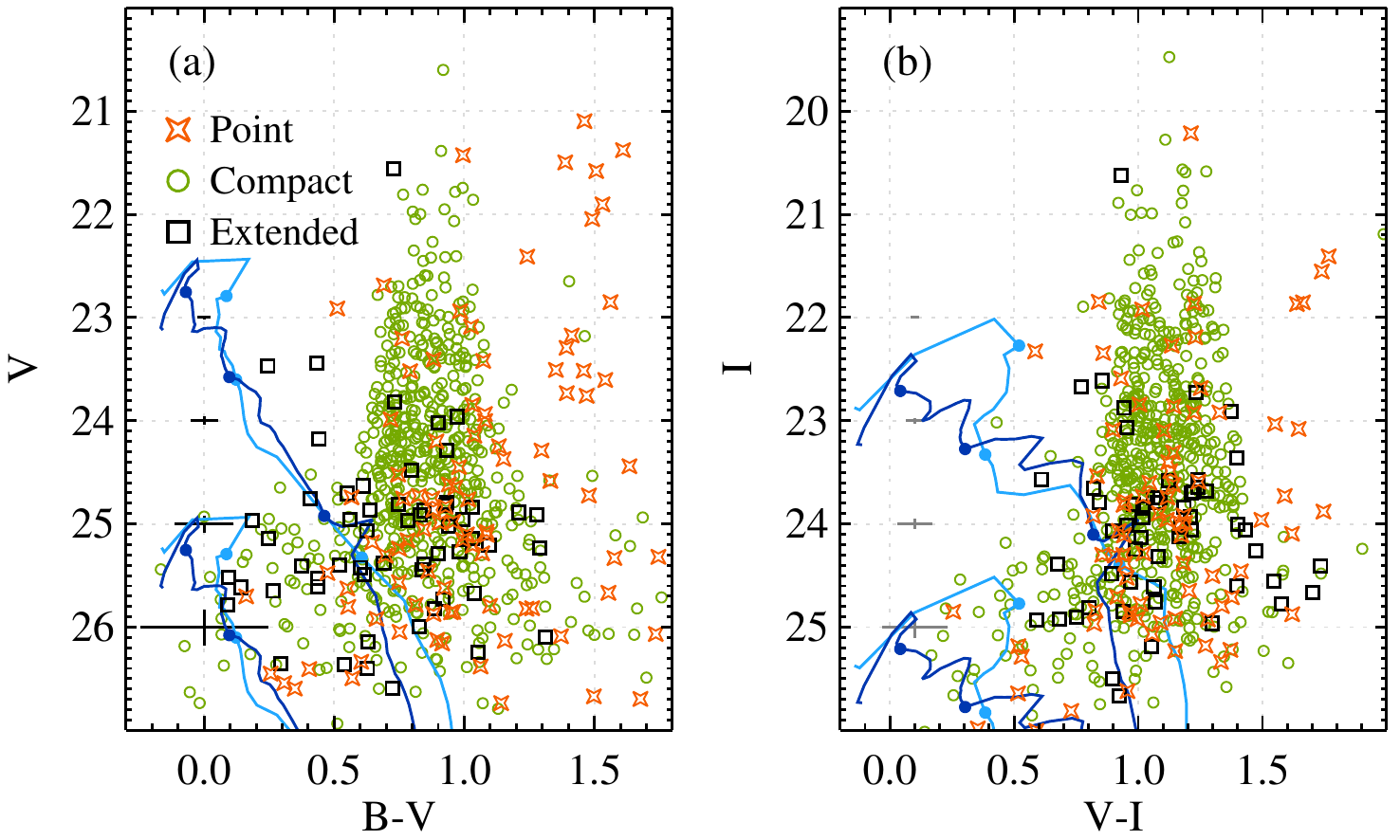}
\caption{$V-(B-V)$ (Left panel) and $I-(V-I)$ (Right panel) CMDs of the selected sources with $C\leq1$ and $V\le27$ mag
in NGC 4589:
The point sources with $C\leq0.52$ (orange starlets),
the compact sources with $0.52<C\leq0.8$ (green circles),
and the extended sources with $0.8<C\leq1.0$ (black boxes). 
The point and compact sources with $0.6<(B-V)\le 1.2$ $(0.8<(V-I)\le 1.4)$ in the vertical plume are mostly globular clusters in NGC 4589. 
The very red point sources at $1.3<(B-V)\le $  1.6 $(1.5<(V-I)\le 1.8)$ are mostly dwarf stars in the Milky Way.
The extended sources are mostly background galaxies.
The mean photometric errors for given magnitude of all the detected sources are plotted. 
Padova SSP models with 
solar (cyan lines) and 0.1 solar (blue lines) metallicities are overplotted.  
Upper and lower SSP lines indicate 
masses $M_{SC} = 10^{4.5}$  and $10^{3.5} M_{\odot}$. 
Ages for $10^7, 10^8$ and $10^9$ years are marked by filled circles on the SSP models. 
}
\label{fig_CMD}
\end{figure*}

\section{Results}

\subsection{Color-magnitude Diagrams of the Star Clusters}

We display the color-magnitude diagrams (CMDs) of the point sources ($C\le 0.52$), the compact sources ($0.52<C\le0.8$), and the extended sources ($0.8 < C\leq1.0$) with $V\le27$ mag
in {\color{blue}\bf Figure  \ref{fig_CMD}}.
The error bars in the left side represent median photometric errors for given magnitudes derived from the photometry of all the measured sources. 
We also plotted Padova simple stellar population (SSP) models for solar metallicity (cyan lines) and $0.1 Z_\odot$ (blue lines) 
with cluster mass $M/M_\odot=10^{3.5}$ and  $10^{4.5}$ \citep{gir00}.
  
These CMDs show several notable features.
First, the most distinguishable feature is a broad vertical branch 
at
$0.6<(B-V)\leq1.2$ ($0.8<(V-I)\leq1.4$), the brightest of which reaches $V\approx 21$ mag ($I\approx 20$ mag).
This branch consists of a dominant population of compact sources and a small number of point sources.
The colors of these sources are similar to those of typical globular clusters. 
Thus the sources in this branch are mostly considered to be globular clusters in NGC 4589.
The branch also contains a small population of extended objects whose colors are numerically not much different from those of the globular clusters. They are mostly background galaxies. 

\begin{figure*} 
\centering
\includegraphics[scale=0.9]{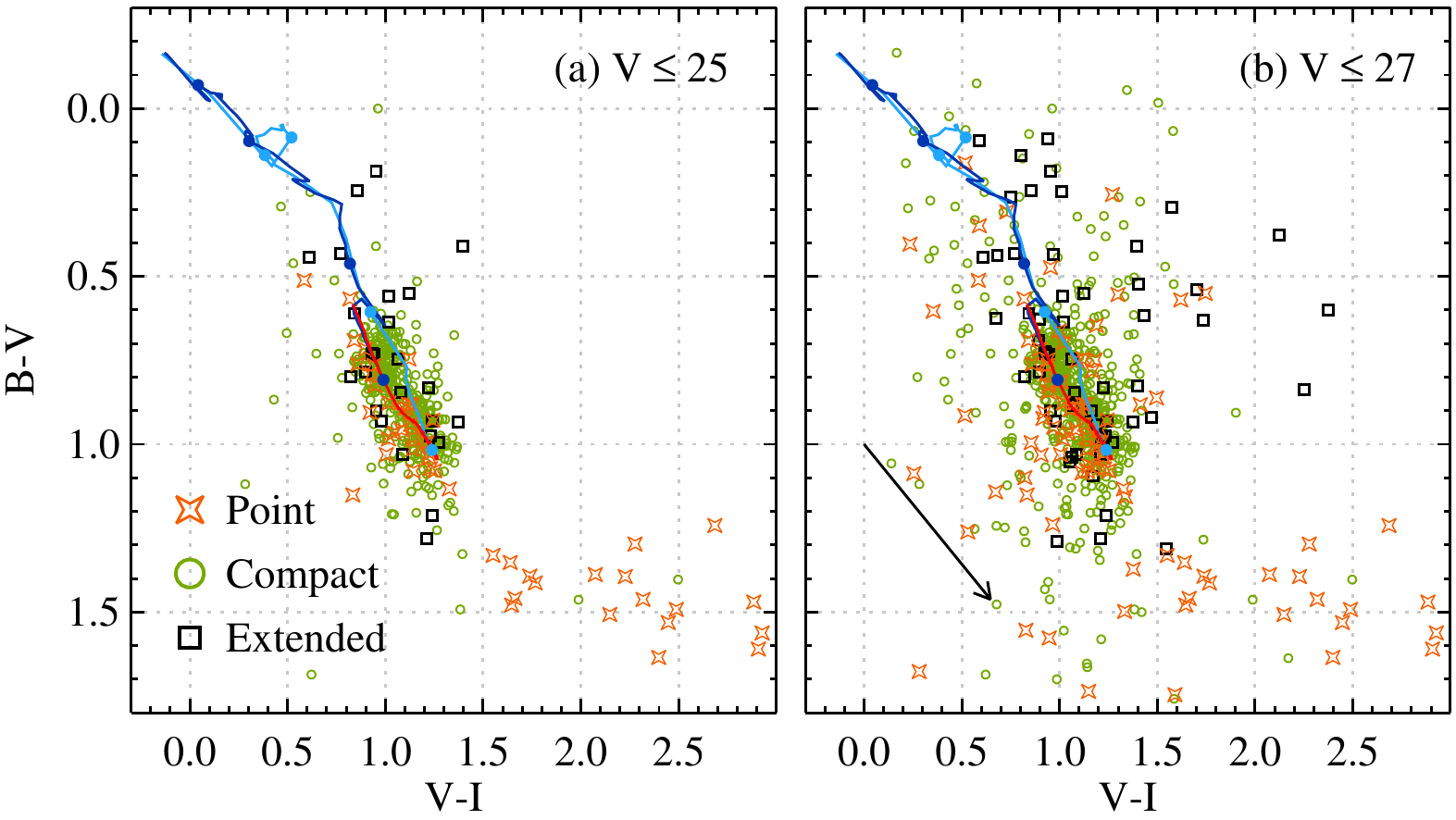}%
\caption{Color-color diagrams of  the selected sources with $V\leq25$ mag 
(a) and $V\leq27$ mag (b) 
in NGC 4589. 
A Padova SSP model for 12 Gyr and [Fe/H] = --2.3 to 0.2 is overplotted by a red solid line in each panel.
Two lines indicate SSP models with solar (cyan) and 0.1 solar (blue) metallicities. 
Filled circles mark Log(age[yr]) = 7, 8, 9, and 10. 
The point and compact sources in the strong concentration area at $0.6<(B-V)\le 1.2$ and $0.8<(V-I)\le 1.4$
are mostly globular clusters in NGC 4589.
The reddening vector is marked by an arrow. 
}
\label{fig_CCD}
\end{figure*}

Second, there is a small population of blue compact sources at 
$(B-V)\leq0.5$ $((V-I)\leq0.7)$, 
which are bluer than the blue limit for the globular clusters. They are mostly fainter than $V=25$ mag ($I=24$ mag) so that they are much fainter than the majority of the globular clusters. These sources can be either young star clusters in NGC 4589 or background galaxies.
To investigate their nature further, we inspect their spatial and radial distributions in the following sections.

Third, 
there is a narrow vertical sequence of very red point sources with $21<V\leq25$ mag at 
$1.3<(B-V)\leq1.6$. 
They have $(V-I)$ colors of 1.5 -- 3.0, as seen in {\color{blue}\bf Figure \ref{fig_CCD}}, 
and many of them present in the $(B-V)$ panel are outside the plotted range for the $(V-I)$ plot. 
They are mainly red dwarf stars in the Milky Way.

Fourth, extended sources show a much broader color distribution and fainter magnitudes compared with compact sources. The number of extended sources is much smaller than that of compact sources, and the spatial distribution of extended sources appears to be uniform.

\begin{deluxetable*}{lcccccccccc}
\setlength{\tabcolsep}{0.05in}
\tablecaption{A Summary of GMM Tests for Color Distributions of the GCs with $V\le25$ mag in NGC 4589}
\tablewidth{0pt}
\tablehead{
& & \multicolumn{3}{c}{Blue GCs} & \multicolumn{3}{c}{Red GCs} \\
& \colhead{Color} &\colhead{Mean} &\colhead{$\sigma$} & \colhead{$N_{\rm total}$} & \colhead{Mean} &\colhead{$\sigma$} & \colhead{$N_{\rm total}$}  & \colhead{$D^c$} & \colhead{$p^c$} & \colhead{$k^c$}
}
\startdata
\multirow{2}{*}{Homoscedastic$^a$} & $(B-V)$ &    $0.78\pm0.01$ &    $0.09\pm0.01$ &    $200\pm21$ &     $0.99\pm0.02$ &     $0.09\pm0.01$ & $138\pm21$ & $2.38\pm0.28$ & $1.94e$--4 & $-0.493$\\ 
& $(V-I)$&    $1.00\pm0.01$& $0.08\pm0.01$& $201\pm14$ &     $1.20\pm0.01$ & $0.08\pm0.01$ & $137\pm14$ & $2.60\pm0.20$ & $3.93e$--6 & $-0.702$ \\
\hline
\multirow{2}{*}{Heteroscedastic$^b$} & $(B-V)$& $0.73\pm0.01$& $0.06\pm0.02$ & $107\pm39$ & $0.93\pm0.03$ &     $0.12\pm0.02$ & $231\pm39$ & $2.06\pm0.38$ & $1.37e$--5 & $-0.493$\\
& $(V-I)$ & $0.99\pm0.02$& $0.07\pm0.01$ & $174\pm33$ & $1.18\pm0.02$ & $0.09\pm0.01$ & $164\pm40$ & $2.50\pm0.24$ & $2.94e$--5 & $-0.702$ \\
\hline
\enddata
\tablenotetext{a}{Same variances for the Gaussian fits.} 
\tablenotetext{b}{Varying variances for the Gaussian fits.}
\tablenotetext{c}{$D$ represents the ratio of the separation of the two peak colors relative to their widths, $p$ denotes the probability for unimodal distribution, and $k$ is the kurtosis.}
\label{tab_cdf}
\end{deluxetable*}

\begin{figure} 
\centering
\includegraphics[scale=0.7]{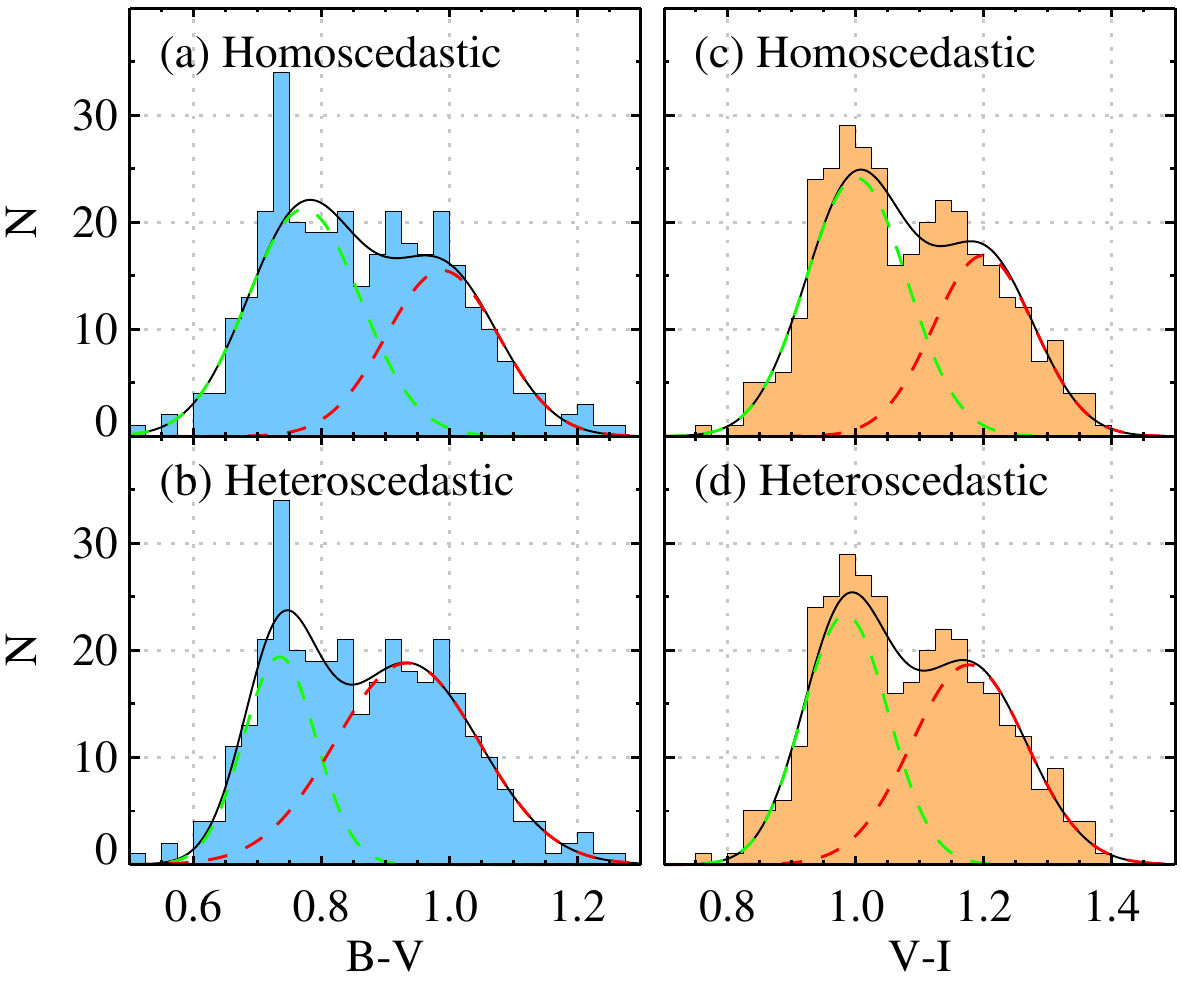}
\caption{$(B-V)$(a,b) and $(V-I)$(c,d) color distributions of the globular cluster candidates ($C\leq0.8$) with $V\leq25$ mag at $0\farcm2<R<2'$ in NGC 4589
for the same variance and the varying variance, respectively.
The sources with $0.6<(B-V)\le 1.2$ ($0.8<(V-I)\le 1.4$)
are mostly globular clusters in NGC 4589.
Solid curved lines denote the fitting with GMM, 
and green and red dashed lines are the Gaussian components for blue GCs and red GCs, respectively. 
}
\label{fig_CDF}
\end{figure}

\subsection{Color-color Diagrams of the Star Clusters}

{\color{blue}\bf Figure \ref{fig_CCD}} shows  the color-color diagram of the the point sources, the compact sources, and the extended sources.
Left and right panels display the sources with $V\leq25$ mag ($M_V\leq-7.1$ mag) and those with $V\leq27$ mag ($M_V\leq-5.1$ mag), respectively.
We also plotted Padova evolutionary tracks
for the SSPs with solar metallicity (cyan lines) and $0.1 Z_\odot$ (blue lines), and the 12 Gyr isochrones for [Fe/H] = --2.3 to +0.0  (red lines).
The {filled circles along the evolutionary track represent the age, 10$^7$, 10$^8$, 10$^9$, and 10$^{10}$ yr.
In this diagram 
the globular clusters are located along the 12 Gyr isochrones for a large range of metallicities, 
showing that they are indeed old globular clusters with a large range of metallicities.

The blue sources with $(B-V)\leq0.5$ (and $(V-I)<1.0$) are located around the SSP models with age $<1$ Gyr, but with a large scatter. 
The mean photometric errors of these blue sources  are 
${\rm err}(B - V)=0.13$ $({\rm err}(V - I)=0.12$) for $25<V\leq26$ mag and ${\rm err}(B - V)=0.25$ $({\rm err}(V - I)=0.21$) for $26.0<V\leq27$ mag.
Therefore, the large scatter in color is mainly due to 
photometric errors. 

There are a small number of extended sources which overlap with the color-color sequence of globular clusters. One of the extended sources has $V=21.6$ mag, 
as bright as the brightest globular clusters, and the rest are more than two magnitudes fainter than this. The effective radius of the brightest extended source is 7.6 pc, 
smaller than the values for ultra compact dwarfs (UCDs), so it is considered to be an extended bright globular cluster. Thus none of them are found to be UCDs. 

\subsection{Color Distributions of the Globular  Clusters}
 
{\color{blue}\bf Figure \ref{fig_CDF}} shows the 
color distributions of the bright star cluster candidates with $C\leq0.8$ and $V\leq25$ mag
 at $0\farcm2<R<2'$ in NGC 4589. 
The most prominent feature in this figure is due to globular clusters in NGC 4589.
The color distributions of these globular clusters show two peaks in both $(B-V)$ and $(V-I)$ colors, clearly suggesting that they are bimodal, as often  seen in other early-type galaxies.

We applied a Gaussian Mixture Modeling (GMM) test \citep{mur10} to this sample. 
We chose an option for the same variance 
(homoscedastic case) as well as the varying variance (heteroscedastic case). 
The results of this test are summarized in {\color{blue}\bf Table \ref{tab_cdf}}.
Key parameters from this test are
the probability for unimodal distribution ($p$), the ratio of the separation of the two peak colors relative to their widths ($D$), and the kurtosis $k$.
If $D$ is larger than two, the distribution is considered to be bimodal. 
A negative value for $k$ is necessary for bimodal distributions.

\begin{figure*} 
\centering
\includegraphics[scale=0.9]{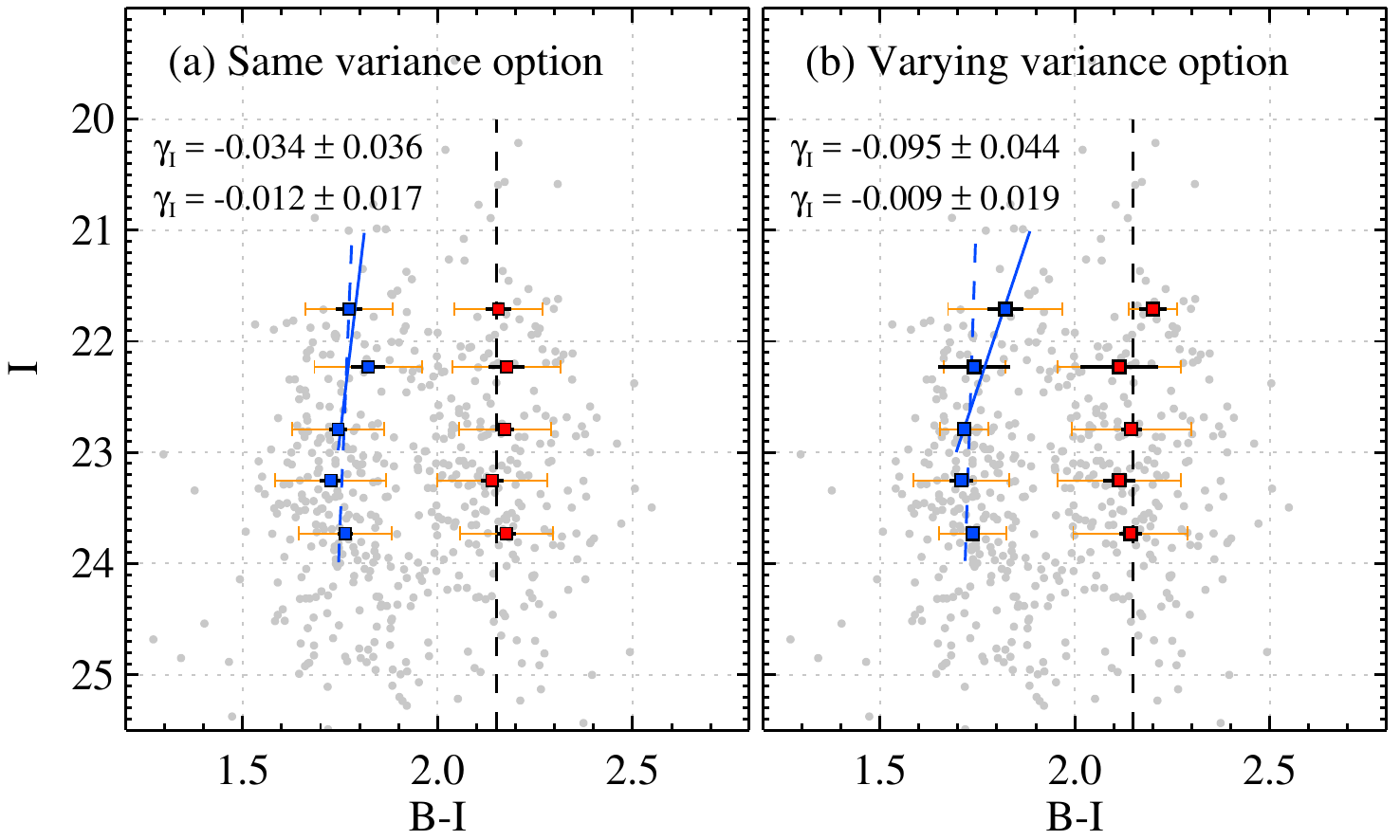}
\caption{$I - (B-I)$ CMDs of the globular cluster candidates ($C \leq 0.8$ and $0.6<B-V\leq 1.2$) in NGC 4589.
The mean colors of the blue and red subpopulations were measured using the GMM code with two options: a same-variance option (a) and a varying-variance option (b).
Yellow and black error bars indicate the sigmas and  errors of the mean colors.
The solid and dashed lines represent linear fits for the blue subpopulations with $21 < I\leq 23$ mag and $21 < I\leq 24$ mag. 
Measured values of the slopes are given in each panel.
The approximate mean color for the red subpopulation is shown by the vertical dashed line.}
\label{fig_BICMD}
\end{figure*}

In {\color{blue}\bf Table \ref{tab_cdf}} the values of $p$ are close to zero ($<10^{-3}$) and the values of $D$ are larger than two for all cases. 
The values of $k$ are smaller than zero in all cases.
Thus, these results show that the color distributions are indeed bimodal.
The blue and red peaks are found to be at  $(B-V)=0.78\pm0.01$ and $0.99\pm0.02$ ($(V-I)=1.00\pm0.01$ and $1.20\pm0.01$) for the homoscedastic case, 
and at $(B-V)=0.73\pm0.01$ and $0.93\pm0.03$ ($(V-I)=0.99\pm0.02$ and $1.18\pm0.02$)  for the heteroscedastic case.  
In the case of the homoscedastic option, the number ratio of the blue GCs and the red GCs derived from the $(B-V)$ colors  (200 vs. 138)  is similar to the value from the $(V-I)$ colors (201 vs. 137).  However, in the case of the heteroscedastic option, the number ratio of the blue GCs and the red GCs derived from the $(B-V)$ colors (107 vs. 231) is significantly different from the value based on the $(V-I)$ colors (174 vs. 164).  Therefore, the results for the homoscedastic option appear to be more reliable.

In summary, we select, as the 
globular clusters, the compact and point sources ($C\le0.8$) with globular cluster-like colors of
$0.6<(B-V)\le1.2$. 
For the following analysis, we divided the globular cluster sample into two subgroups according to their color: 
blue (metal-poor) globular clusters with $0.6<(B-V)\leq 0.85$ and  red (metal-rich) globular clusters with $0.85<(B-V)\leq 1.2$. 
We chose, as the division colors,
the colors with a minimum value between the two peaks in the color histograms.

Blue globular clusters in brightest cluster galaxies often show a color-magnitude relation 
known as the blue-tilt (see
\citep{har06,har17} and references therein). 
In {\color{blue}\bf Figure \ref{fig_BICMD}}, we display the $I-(B-I)$ CMD of the globular clusters in NGC 4589 to check the presence of any blue tilt. 
We divided the bright globular clusters with $21<I\leq24$ mag into five groups according to their $I$-band magnitudes in steps of $\Delta I = 0.5$ mag. We set the brightest magnitude bin to sample all the globular clusters with $21 < I\leq22$ mag.
The number of globular clusters in these  groups ranges from 46 ($21 < I\leq22$ mag) to 108 ($23.0 <I\leq23.5$ mag).

We measured the mean colors of the blue and red globular clusters in each group using the GMM code.
The measured values of $D$ were larger than 2 in all groups, but the values of $p$ were not always close to zero ($<10^{-3}$). 
This indicates that  the unimodal distribution is not always rejected,  although the bimodal distribution is meaningful in all cases.
The same-variance and varying-variance options in the code yield similar colors for the faint bins ($I<22.5$ mag), but slightly different values for the brighter bins ($I>22.5$ mag).
The difference in the brighter bins is due to the smaller sample size compared to that of the fainter bins.

\begin{figure*} 
\centering
\includegraphics[scale=0.8]{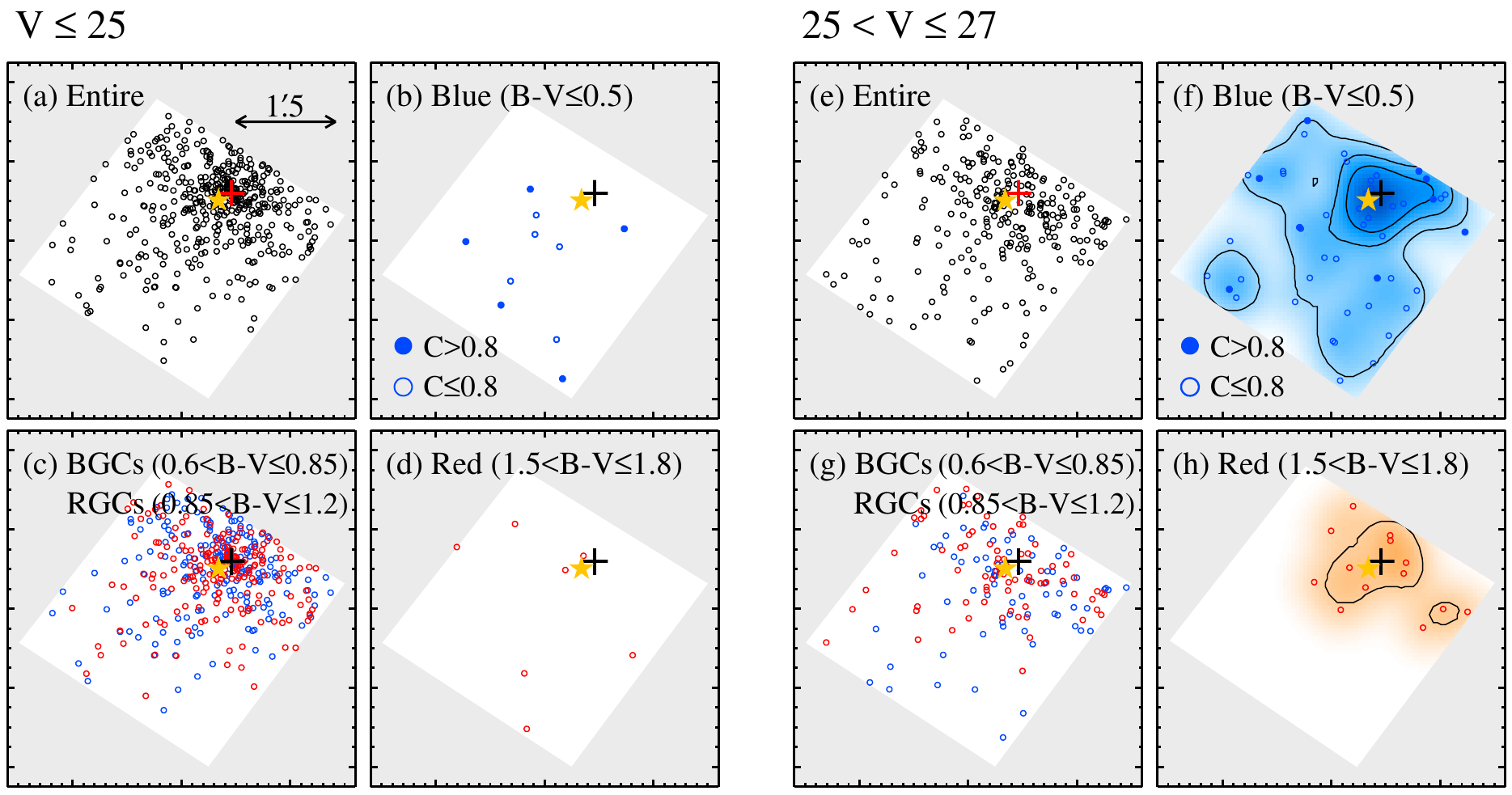}
\caption{Spatial distributions of the selected point and compact sources ($C\leq0.8$) 
with $V\leq25$ mag (left) and  $25<V\leq27$ mag (right) in NGC 4589:
(a,e) all sources, 
(b,f) blue sources ($(B-V)\leq0.5$), 
(c,g) blue GCs and red GCs ($0.60<(B-V)\le 0.85$ and
$0.85<(B-V)\le 1.20$),
(d,h) red sources ($1.5<(B-V)\le 1.8$).
In (b) and (f) we also plotted the extended blue sources ($C>0.8$) for comparison.
Contours in (f) and (h) represent the number density map of the blue sources and the red sources with $25<V\leq27$ mag.
The locations of the galaxy center and SN are marked by the cross and the yellow starlet. 
}
\label{fig_spat}
\end{figure*}

Fitting the bright blue globular clusters with $21 < I\leq23$ ($-11.1 < M_I\leq-9.1$) mag, we obtained the values of the slope, $\gamma_I (21<I\leq23)= d(B-I)/dI = -0.034\pm0.036$ for the homoscedastic option and $\gamma_I (21<I\leq23)= -0.095\pm0.044$ for the heteroscedastic option.
If we extend the magnitude range down to $I = 24$ ($M_I=-8.1$) mag, we obtain the slope values, 
$\gamma_I (21<I\leq24)= -0.012\pm0.017$ and $-0.009\pm0.019$ for the homoscedastic and heteroscedastic options, respectively.
Thus, the slope value for the bright globular cluster sample ($21 < I\leq23$ mag) derived with the heteroscedastic option shows a hint of blue tilt at the level of $2\sigma$, while the values for other cases do not. 

\citet{har06} presented an $M_I - (B-I)_0$ CMD for the combined sample of globular clusters in eight brightest cluster galaxies (BCGs) (their Figure 21), which shows clearly a blue tilt for $-11.8<M_I<-9.5$ mag.
They provided only a mass ($M$)$-$metallicity ($Z$) relation derived from the CMDs, $Z \propto M^{0.55}$, and did not present the value of the blue tilt slope in the CMD.
We estimate the value of the slope for the blue tilt in their Figure 21, obtainging
$\gamma_I \approx  -0.09$. 
Thus the slope value for the bright globular cluster sample ($21 < I\leq23$ mag) of NGC 4589 derived with the heteroscedastic option in this study,  $\gamma_I = -0.095\pm0.044$,  is similar to the mean slope for the BCGs in \citet{har06}.
Recently \citet{har17} presented an $M_{F814W} - (F475W-F814W)_0$ CMD for the combined sample of globular clusters in other five BCGs (see their Figure 22), and they pointed out that the estimated slopes of the blue tilt show a large spread among galaxies, which range from $\gamma_M =dLogZ/dLogM \sim 0$ to +0.27 
(or $\gamma_I =d(F475W-F814W)/dF814W \sim 0$ to --0.05). These values are much smaller than that given in \citet{har06}, $\gamma_M =0.55$.

\subsection{Spatial Distributions of the Star Clusters}

In {\color{blue}\bf Figure \ref{fig_spat}}
we plotted the spatial distributions of the point and compact sources 
with $V\leq25$ mag (left panels) and  $25<V\leq27$ mag (right panels): (a)(e) all sources,
(b)(f) blue sources ($(B-V)\le0.5$, bluer than the globular clusters),
(c)(g) blue globular clusters ($0.6<(B-V)\le 0.85$) and red globular clusters ($0.85<(B-V)\le 1.2$), and
(d)(h) red sources 
($1.5<(B-V)\le 1.8$), redder than the globular clusters. 
In {\color{blue}\bf Figures \ref{fig_spat}(b) and (f)}, we also plotted the spatial distributions of the  extended blue sources with $C>0.8$ with filled circles in order to check their membership. The spatial distributions  of these extended sources do not show any central concentration, which indicates that they are background galaxies. 
We also marked the positions of the galaxy center (cross) and SN 2005cz (yellow starlet) in the figure.

A few interesting features are noted in this figure.
First, the spatial distributions of both blue and red globular clusters in {\color{blue}\bf Figures \ref{fig_spat}(c) and (g)} show a strong central concentration around the galaxy center.
This result implies that these sources are indeed globular clusters that are gravitationally bound to NGC 4589.

Second, the spatial distribution of the bright red sources (with $V\le25$ mag) in {\color{blue}\bf Figure \ref{fig_spat}(d)}  is roughly uniform, which indicates that they are not the members of NGC 4589. Considering that the majority of these sources are point sources,  
we conclude that they are red dwarf stars in the Milky Way. 
%

Third, the number density contour maps in {\color{blue}\bf Figure \ref{fig_spat}(f)}  indicate that the spatial distribution of the faint blue sources appears to show a weak central concentration around the galaxy center.  The nature of the faint blue sources will be checked further with their radial number density profile in the following section. 

Fourth, the number density contour map in {\color{blue}\bf Figure \ref{fig_spat}(h)} indicates that the spatial distribution of the faint red sources appears to show a very weak  central concentration around the galaxy center, but the number of the sources is too small to be significant. 
In this figure, we plotted the red sources $1.5<(B-V)\le 1.8$, to avoid any contamination due to the faint red globular clusters. 

\begin{figure} 
\centering
\includegraphics[scale=0.9]{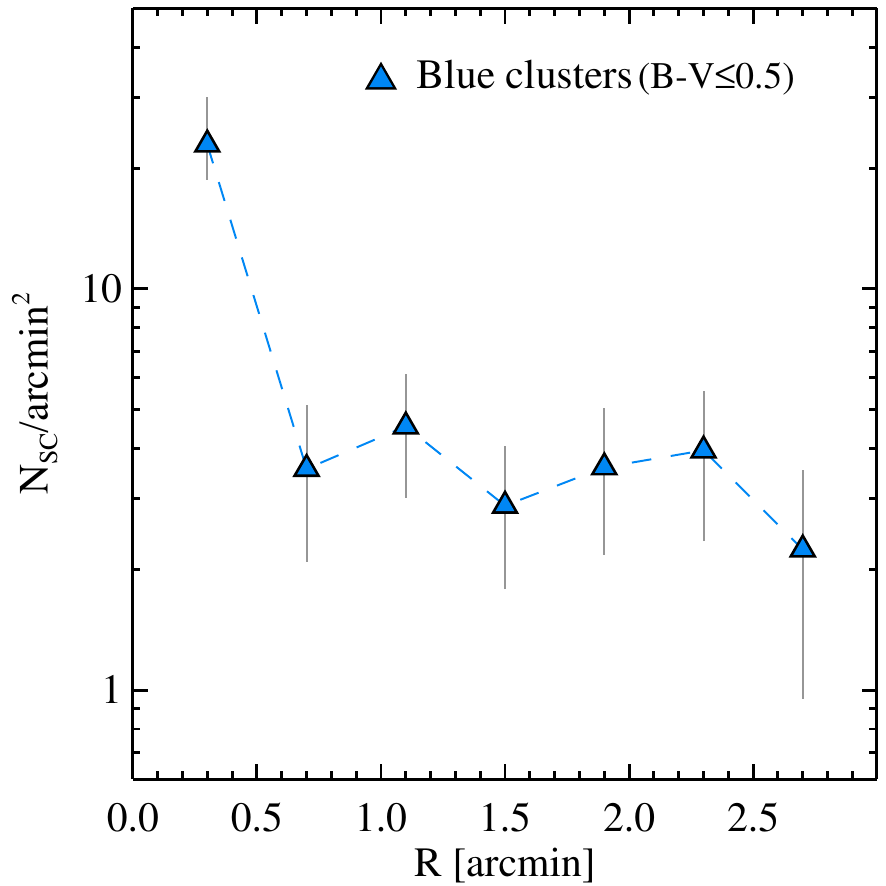}
\caption{
Completeness-corrected radial number density profile for the blue ($B-V\leq0.5$)
sources with $V\leq27.0$ mag in NGC 4589.
The central region at $R<0\farcm1$ is masked out.
Note that the blue 
sources show a clear central excess at  $R\lesssim0\farcm5$,
which indicates that they are young star clusters in NGC 4589. 
}
\label{fig_RDPblue}
\end{figure}

\subsection{Radial Distributions of the Young Star Clusters}

To further investigate the nature of the blue sources seen in the CMDs, we selected the blue sources with $C\le0.8$, $(B-V)\le0.5$, and $V\le27.0$ mag. 
The number of these blue sources is 46. 
Then we derived their radial number density profile, as plotted in 
{\color{blue}\bf Figure \ref{fig_RDPblue}}. 
The errors for the number density are Poisson errors. 

 In deriving the radial density profiles, we excluded the central region at $R<0\farcm1$, which masked out most of the dust lane.
We also excluded out-of-the-edge/gap regions.
We calculated the true area by counting individual ``pixels'' 
which were used for the calculation of cluster properties. 
Pixels around saturated stars, background galaxies, and masked regions were not used.
We corrected the incompleteness of our photometry using the completeness test results as a function of galactocentric distance. 

The radial profile of the blue sources  clearly  shows a central excess at $R<0\farcm5$ ($\sim$3.8 kpc) and a flattening at the level of $\Sigma \approx 3$ arcmin$^{-2}$ in the outer region at $R>0\farcm5$. 
This result shows that the blue sources in the central region at $R<0\farcm5$ are mainly the members of NGC 4589, while those in the outer region at $R>0\farcm5$ are 
most likely background galaxies. Considering as well that they are mostly compact (slightly resolved) sources, we conclude that the blue sources in the central region are mainly genuine young star clusters in this galaxy. 

We estimate the background level for the blue sources using the region at $R>1'$,
and we obtain  $\Sigma =3.0\pm1.3$ arcmin$^{-2}$. 
Subtracting the background level, we derive the total number of the young star clusters with $24.5<V\leq27$ at $R<0\farcm5$  and obtain $N_{\rm YSC}=16\pm6$.
From the comparison of these sources with the SSP models for solar metallicity and $Z>0.1Z_\odot$ \citep{gir00} as shown in {\color{blue}\bf Figure \ref{fig_CMD}}, we estimate their ages to be 10 Myr -- 1 Gyr and their stellar masses 
to be $M=10^{3.5}$ to $10^{4.5} M_\odot$. 
These values change little even if we use the SSP models for $Z>0.1Z_\odot$. 

\subsection{Radial Distributions of the Globular Clusters}
We derived the radial number density profiles of the bright globular clusters with $V\le25$ mag in NGC 4589:
all globular clusters,
the blue globular clusters, and
the red globular clusters.
We corrected the radial profiles for the incompleteness using the results from the artificial source test.

Since the $HST$ field is not large enough to cover the background region, it is difficult to estimate the background level.
Therefore, we present the results without background correction. The results for the outer region at $R>2'$ are uncertain.
{\color{blue}\bf Figure \ref{fig_rad}(a)} displays the radial number density profiles of these globular clusters.

Fitting the profiles for $R\leq2\farcm5$ with a S{\'e}rsic law \citep{ser63}, we obtain the S{\'e}rsic index values, 
$n_{\rm AGC} = 1.41\pm 0.34$ for all globular clusters, 
$n_{\rm BGC} = 1.07\pm 0.28$ for the blue globular clusters, and
$n_{\rm RGC} = 1.89\pm 0.83$ for the red globular clusters.
These values are similar to those of NGC 4921 \citep{lee16}.
The corresponding effective radii of the globular cluster systems are
$R_{\rm eff, AGC}=0\farcm94\pm0.09$, 
$R_{\rm eff, BGC}=0\farcm90\pm0.08$, and
$R_{\rm eff, RGC}=1\farcm00\pm0.22$, respectively.
Although the S{\'e}rsic index value for the blue globular clusters is slightly smaller than that for the red globular clusters, the difference is within errors. 
In most massive galaxies, the blue globular clusters have a shallower density profile than the red globular clusters(\citet{lee98,bro06} and references therein; see also \citet{lee16,cho16,har17}). Therefore, NGC 4589 is a rare example that shows both blue  and red subpopulations that have similar radial profiles. 

For comparison we also plotted the $V$-band radial surface brightness profile, $\mu_V$, of the galaxy light derived from the F555W image using the IRAF/ellipse task. We estimated the background levels using the area at $R>2\farcm5$, and used them for  background subtraction.

\begin{figure} 
\centering
\includegraphics[scale=0.9]{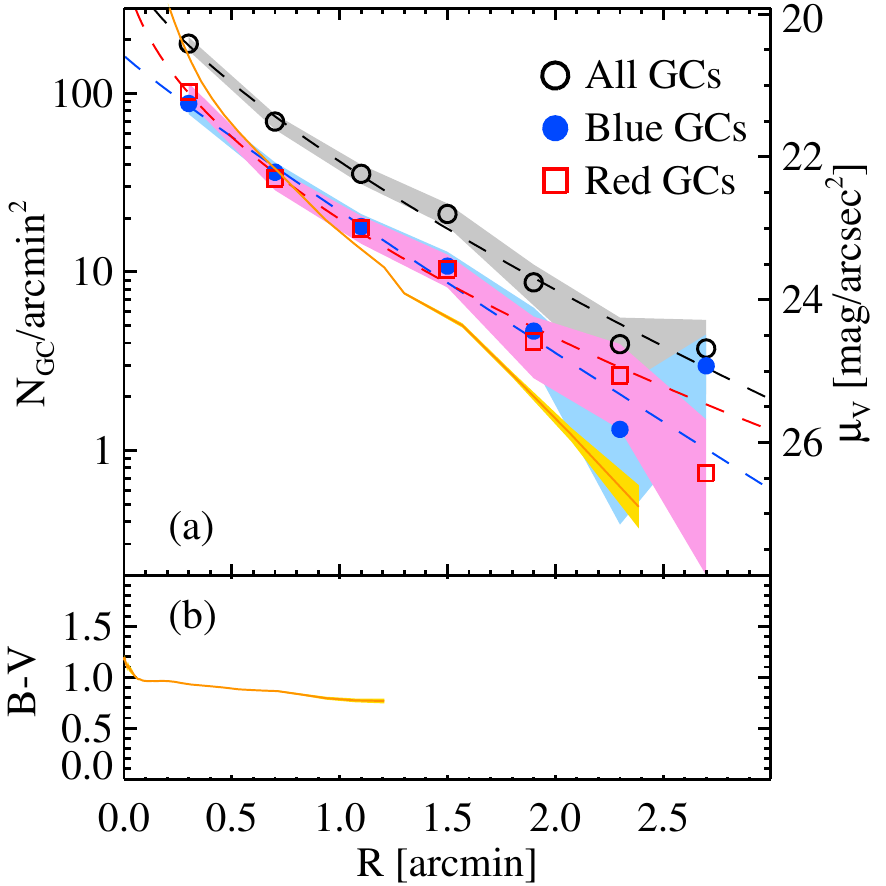}
\caption{(a) Radial number density profiles of 
all GCs (black open circles), blue GCs  filled circles), and red GCs (red squares) with 
$V\leq25.0$ mag in NGC 4589. Shaded regions represent the errors for the number density.
The yellow solid line represents the $V$-band surface brightness profile 
 of NGC 4589. 
Its magnitude scale is indicated in the right $Y-$axis. 
Dashed lines denote the S{\'e}rsic profile fitting to the data for 
all GCs ($n_{\rm AGC}=1.41\pm0.34$), 
blue GCs ($n_{\rm BGC}=1.07\pm0.28$), and 
red GCs ($n_{\rm RGC}=1.89\pm0.83$).
(b) The radial profile of the $(B-V)$ color of the galaxy light. Shaded regions denote the photometric error of the colors.
}
\label{fig_rad}
\end{figure}

We converted the unit for the surface brightness profile of the galaxy light, the magnitude per square arcsec, into Log (flux per square arcmin).  
In the figure we plotted $\mu_V/2.5$ for direct comparison with the radial number density profiles of globular clusters. 
We shifted the resulting radial profile of the galaxy light 
along the vertical direction to match the radial number density profile of the blue and red globular clusters at $R\sim0\farcm7$. 
In the right y-axis we labeled the scale for the surface brightness profile. 

S{\'e}rsic law fitting for the galaxy light with $0\farcm1 < R < 2\farcm0$ gives a S{\'e}rsic index of 
$n_{\rm galaxy}=3.11\pm0.21$ (with $R_{\rm eff, {\rm galaxy~light}}=0\farcm47\pm0.01$), 
showing that the galaxy light follows 
 the de Vaucouleurs law. 
The radial profile of all globular clusters is flatter than the surface brightness profile of
the galaxy light, and the effective radius of the entire globular cluster system ($R_{\rm eff, AGC}=0\farcm94\pm0.09$) is two times larger than the value for the galaxy light ($R_{\rm eff, {\rm galaxy~light}}=0\farcm47\pm0.01$). 
This shows that the globular cluster system is more extended than the galaxy light, which is also seen in many other galaxies (see \citet{bro06} and references therein).

In {\color{blue}\bf Figure \ref{fig_rad}(b)} we plot the radial variation of the $(B-V)$ color 
of the galaxy light. 
The color 
of the galaxy light} is redder than $(B-V)=1.0$ in the central region at $R\leq0\farcm1$, and becomes slowly bluer from $(B-V)=1.0$ at $R=0\farcm1$ to $(B-V)=0.8$ at $R=1\farcm2$.
As shown in the color-color diagram ({\color{blue}\bf Figure \ref{fig_CCD}}), SSP models show that an old stellar population with $-2.3<$[Fe/H]$<+0.2$ can have $(B-V)$ colors ranging  from 0.6 mag to 1.1 mag. Thus, the integrated color of the galaxy light is consistent with colors of the old stellar populations with a large range of metallicities. 

\begin{figure*}
\centering
\includegraphics[scale=1.1]{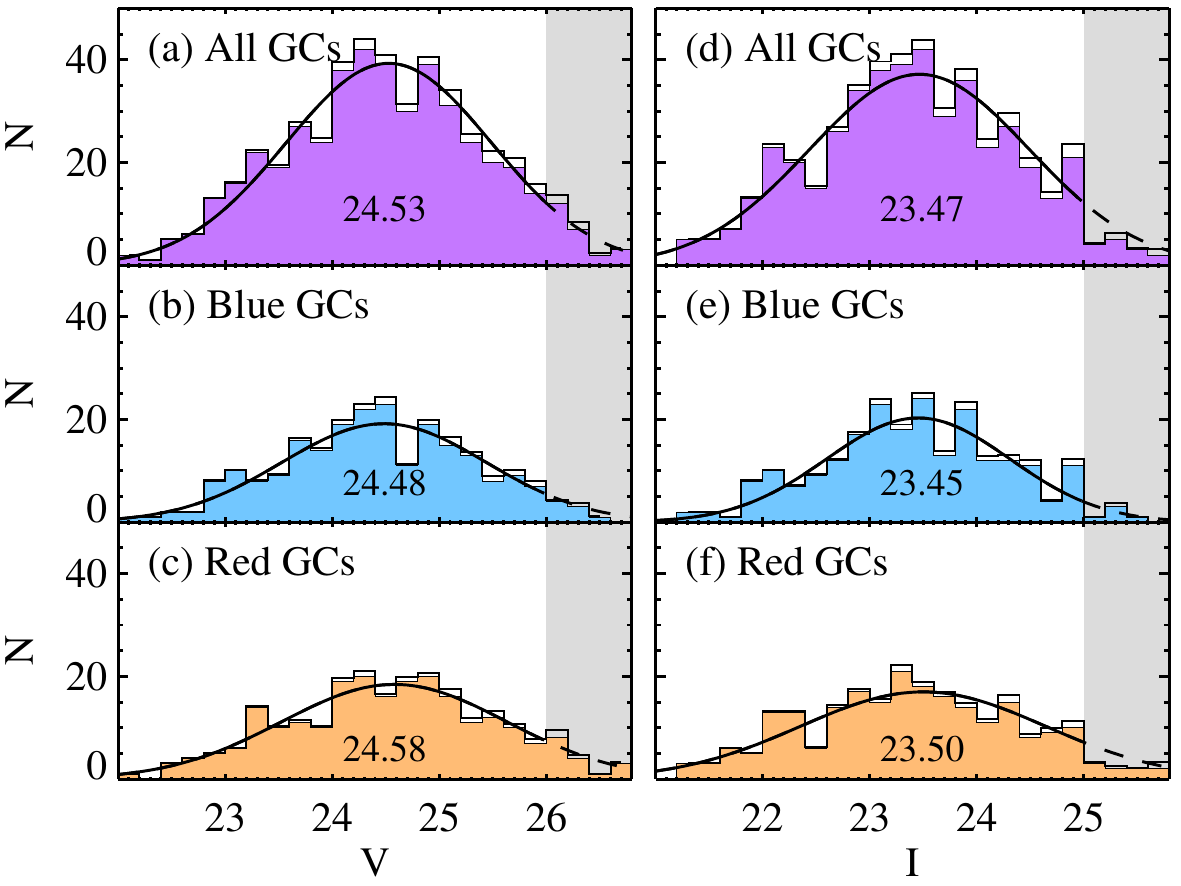}
\caption{
$V$-band (left) and $I$-band (right) luminosity functions of all GCs, blue GCs, and red GCs at $0\farcm2<R\leq2'$ of  NGC 4589. 
 Filled and blank histograms represent the luminosity functions before and after completeness correction. 
Solid lines represent the Gaussian function fitting to the completeness-corrected data
 for $V\leq26.0$ mag, and $I\leq25.0$ mag. 
The numbers in each panel denote the turnover magnitudes.
\label{fig_LF}}
\end{figure*}

\subsection{Luminosity Functions of the Globular Clusters}

We  derived $V$ and $I$-band 
GCLFs for NGC 4589 by counting the number of the globular clusters at $0\farcm2<R\leq2\farcm0$. We excluded the central region at $R<0\farcm2$ where the completeness is much lower than the outer region.
We corrected the GCLFs for the completeness using the results of the completeness test.
In {\color{blue}\bf Figure \ref{fig_LF}} 
we display $V$ and $I$-band GCLFs: for all globular clusters (top panels),
 the blue globular clusters (middle panels), and the red globular clusters (bottom panels).
The GCLFs in the figure appear to be approximately Gaussian, showing a turnover (peak) at $V \approx 24.5$ mag ($I \approx 23.5$ mag).

\begin{deluxetable*}{lcccccc}
\tabletypesize{\footnotesize}
\setlength{\tabcolsep}{0.05in}
\tablecaption{A Summary of Gaussian Fits for LFs of All GCs, Blue GCs and Red GCs in NGC 4589} 
\tablewidth{0pt}
\tablehead{ \multirow{2}{*}{ } &  \multicolumn{3}{c}{$V$} & \multicolumn{3}{c}{$I$} \\
  & \colhead{Center} &\colhead{Width} & \colhead{$N_{\rm total}^a$} & \colhead{Center} &\colhead{Width} & \colhead{$N_{\rm total}^a$}
}
\startdata
All GCs, $0\farcm2 < R \leq 2\arcmin$ 	& $24.53\pm0.06$ & $0.96\pm0.05$ & $472\pm25$  & $23.47\pm0.07$ & $1.03\pm0.07$ & $478\pm26$\\
Blue GCs & $24.48\pm0.09$ & $0.96\pm0.07$ & $230\pm17$  & $23.45\pm0.07$ & $0.86\pm0.06$ & $218\pm16$	\\
Red GCs	 & $24.58\pm0.10$ & $1.05\pm0.10$ & $242\pm19$  & $23.50\pm0.12$ & $1.14\pm0.12$ & $243\pm21$ 	\\
\hline
\enddata
\tablenotetext{a}{The errors for the numbers are from the fitting errors for the integration of Gaussian functions provided by IDL/mpfitexpr.}
\label{tab_gclf}
\end{deluxetable*}

We fit the GCLFs with a Gaussian function for the magnitude range  $V\leq26.0$ mag ($I\leq25.0$ mag) where incompleteness of our photometry is not significant. We obtain the values of the turnover magnitudes and widths as summarized in {\color{blue}\bf Table \ref{tab_gclf}}.
The turnover magnitudes and widths for all globular clusters are
 $V{\rm (max)} = 24.53\pm0.06$ mag with $\sigma = 0.96\pm0.05$ 
and $I{\rm (max)} = 23.47\pm0.07$ mag with $\sigma = 1.03\pm0.07$.

It is noted that the widths of the GCLFs for NGC 4589 are smaller than the value for  Milky Way  globular clusters, $\sigma = 1.2-1.4$ for $V$-band \citep{dic06,rej12}, 
and those for other elliptical galaxies, $\sigma \approx 1.3$, in \citet{kun01}. 
\citet{kun01} could not determine the value of the width for NGC 4589 because of the shallow photometric limit. 
However, the values in this study are similar to those for NGC 4921, the brightest spiral galaxy in Coma, $\sigma = 1.0-1.1$ for $V$ and $I$ bands based on deep $HST$ images \citep{lee16}. 
We note that the data in this study goes much deeper than that in \citet{kun01} and that the size of the globular cluster sample in NGC 4589 is much larger than the size of the Milky Way globular cluster sample. 

\citet{jor07} derived a relation between the $g$ and $z$ band GCLF width and the absolute $B$-band magnitude of their host galaxies from the ACSVCS sample: $\sigma_g =(1.14\pm0.01)-(0.100\pm0.007)(M_{B,{\rm gal}} + 20)$ (similar results, but for $M_{z,{\rm gal}}$,  were given later for the combined sample of ACSVCS and ACSFCS by \citet{vil10}).
If we use this relation for the luminosity of  NGC 4589 ($M_B=-20.47\pm0.23$ mag), we obtain a value of
$\sigma_g =1.19$. The scatter of this relation for the magnitude range of NGC 4589 in Figure 9 of \citet{jor07} is estimated to be about $\pm0.1$.
Thus the GCLF width of NGC 4589 in this study is $\sim$0.2 smaller than the value expected from \citet{jor07} at the level of $2\sigma$.

The $V$-band turnover magnitude
for the blue globular clusters 
($V{\rm (max)} = 24.48\pm0.09$ mag) 
is 0.10 mag brighter than that for the red globular clusters 
($V{\rm (max)} = 24.58\pm0.10$ mag).
However, this difference is at the level of $1\sigma$, so it is not significant. 
On the other hand, the $I$-band turnover magnitude
for the blue globular clusters 
($I{\rm (max)} = 23.45\pm0.07$ mag)
is nearly the same as the value for the red globular clusters 
($I{\rm (max)} = 23.50\pm0.12$ mag).
The widths for the blue globular clusters and the red globular clusters are found to be similar.

\subsection{ The Total Number and Specific Frequency of the Globular Clusters}

We estimate the total number of globular clusters in NGC 4589 using the radial number density profile and luminosity function of the globular clusters, 
and the area coverage of the $HST$/ACS field in mind.
We counted the number of the blue and red globular clusters brighter than V = 25 mag in $R \leq 0\farcm94\pm0.09$, which is the half number radius of all globular clusters we measured from the radial number density profile. 
It is not easy to estimate the foreground/background contamination in this value when considering the limit of the $HST$ field. However, the globular cluster density is moderately high inside the half number radius of the globular cluster system so that the contribution of the foreground/background sources must be not significant.
Assuming a Gaussian luminosity distribution of the globular clusters with a peak at V = $24.53\pm0.06$ mag and a width of $0.96\pm0.05$, we obtain the total number of globular clusters in NGC 4589 to be N(total) = $640\pm50$.
The error is derived from individual measurement errors associated with the half number radius 
and the luminosity distribution 
of the globular clusters. 
From this number and the absolute magnitude of NGC 4589 ($M_V = -21.41\pm0.23$ mag), we estimate the
specific frequency, and obtain $S_N=1.7\pm0.2$. 
This value is similar to the mean value for bright E galaxies with similar luminosities to that of NGC 4589 \citep{har13}.

\subsection{GCLF Turnover Magnitude Calibration}

We estimate the distance to NGC 4589 using the GCLF method as applied to the case of Coma galaxies in \citet{lee16}.
The turnover magnitude of the GCLFs has been used as a standard candle for a long time
\citep{har01,ric03,dic06,rej12}. 
It is known that the $V$-band turnover magnitudes of the metal-poor globular clusters are brighter than or equal to those of the metal-rich globular clusters, and that this magnitude difference between the two subpopulations shows a large spread among 
galaxy samples 
\citep{dic06,rej12}.
   
In an extensive review of the GCLF method,  \citet{rej12} concluded that the luminosity functions of the metal-poor (blue) globular clusters are a better distance indicator than those 
using the entire sample of globular clusters, because they show smaller scatter and because they are independent of the fraction of metal-rich globular clusters which varies depending on the galaxy.
Therefore we use the luminosity functions of the blue globular clusters to estimate the distance to NGC 4589.

We adopt the calibration of the $V$-band  turnover magnitudes for the metal-poor globular clusters given by \citet{dic06}.
For the zero-point calibration of the $M_V$(RR)--[Fe/H] relation to be used for distance estimation of the globular clusters, \citet{dic06} adopted a distance to the LMC $(m-M)_0 = 18.50$. This value is close to the recent geometric measurement based on eclipsing binaries of the LMC \citep{pie13},   
$(m-M)_0= 18.494\pm0.008({\rm ran})\pm0.048({\rm sys})$.
A better calibration of the GCLF with updated data for the GCLF and the distance to their host galaxies is needed in the future. 

\citet{dic06} derived a calibration 
from the selected sample of 74 metal-poor ([Fe/H]$<-1.0$) globular clusters 
located in the outer region at $2<R_{GC}<35$ kpc 
with relatively lower reddening values with $E(B-V)<1.0$ in the Milky Way. 
In the Milky Way, we can observe globular clusters which are even  located close to the galaxy center in addition to the halo globular clusters, but we can only observe mostly halo globular clusters in other galaxies. 
Therefore, the sample of globular clusters we observe in other galaxies is mostly made up of halo globular clusters and it is closer to the sample of MW halo globular clusters, rather than to the entire  sample of   MW globular clusters. 
From this sample \citet{dic06} obtained 
$M_{V,MWG} {\rm (max)} = -7.66\pm0.11$ mag.
Then they derive similar calibrations from the samples of the metal-poor globular clusters in M31 and a set of 14 early-type galaxies in the literature: $M_{V,M31} {\rm (max)} = -7.65\pm0.19$ mag, and $M_{V, ETG} {\rm (max)} = -7.67\pm0.23$ mag. 
These three values 
are in excellent agreement, but they are derived from galaxies with a wide range of morphological types, luminosity, and mass. 
\citet{dic06} suggested, as a final calibration for the metal-poor globular clusters, a weighted mean of these three values,
$M_V {\rm (max)} = -7.66\pm0.09$ mag.

The error of the zero-point for the $M_V(RR)$--[Fe/H] relations adopted for the calibration of the turnover magnitudes in \citet{dic06} is $\pm0.05$ mag. 
Combining this zero-point error and the mean error of the $M_V {\rm (max)}$ calibration, we estimate the systematic error for the metal-poor globular cluster calibration of $M_V {\rm (max)}$ to be $\pm0.10$ mag. 

\begin{deluxetable*}{lcc}
\setlength{\tabcolsep}{0.05in}
\tablecaption{A Summary for GCLF Distance Estimation for NGC 4589}
\tablewidth{0pt}
\tablehead{ \colhead{Parameter} & \colhead{Value} & \colhead{Remarks}}
\startdata
Systematic Errors of the GCLF method && \\
\hline
$M_V({\rm max})$(metal-poor GC) & $-7.66\pm0.09^a$  
 &   \citet{dic06} \\
Zero-point error of $M_V-$[Fe/H](RR) relation &$\pm0.05^b$  
 &   \citet{dic06} \\
Adopted error for metal-poor GC calibration of $M_V$(max)	& $\pm0.10^c$ & This study \\
Intrinsic uncertainty of the turnover magnitudes (metal-poor GC) 	& $\pm0.1^d$ & 
This study \\
Total systematic error of the GCLF method based on metal-poor GCs 	& $\pm0.14^e$ & This study  \\ 
\hline
NGC 4589& &  \\
\hline
Foreground extinction, $A_{V}$ &  $0.077\pm0.04$  & \citet{sch11} \\
V(max)(blue GC) & $24.48\pm0.09$ & This study \\
V(max)$_0$(blue GC) & $24.40\pm0.10$ & after extinction correction\\
Aperture correction error & $\pm0.03$ & This study \\
Transformation error & $\pm0.03$ & \citet{sir05} \\
Total systematic error of the distance modulus & $\pm0.15^f$ & This study \\
 $(m-M)_0$ & $32.06\pm0.18$ 
 & $\pm0.10{\rm (ran)}\pm0.15{\rm (sys)}$\\
Distance, d [Mpc] 	& $25.8\pm2.2$ 
& $\pm1.2{\rm (ran)}\pm1.8{\rm (sys)}$\\
\hline
\enddata
\label{tab_gclfdistance}
\tablenotetext{a}{ 
 Based on the samples of the metal-poor globular clusters in the Milky Way, M31 and 14 early-type galaxies \citep{dic06}.
The error denotes the mean error of the GCLF turnover magnitudes for the three samples of calibrator galaxies.} 
\tablenotetext{b}{The error of the zero-point for the $M_V(RR)-$[Fe/H] relations used in \citet{dic06}.} 
\tablenotetext{c}{The sum of the mean calibration error and the error for the $M_V(RR)-$[Fe/H] relations.}
\tablenotetext{d}{Note that intrinsic uncertainty of the turnover magnitudes based on the entire globular clusters is $\pm0.2$  mag \citep{ric03,rej12}.}
\tablenotetext{e}{The sum of the calibration error  of $M_V$(max) and intrinsic uncertainty of the turnover magnitudes of the metal-poor globular clusters.}
\tablenotetext{f}{The sum of  the aperture correction and transformation errors for NGC 4589 and  the total systematic error of the GCLF method based on metal-poor globular clusters.}
\end{deluxetable*}

There is another source for the systematic error of the GCLF method, which is an intrinsic uncertainty 
(scatter) of the turnover magnitudes of the GCLFs. 
In the review of the GCLF method, 
\citet{rej12} (as well as Rejkuba (2018, private communication)) discussed several factors for the errors of the GCLF methods that use the entire globular cluster sample in a galaxy, which include environment effect, dependence on the properties of their host galaxies, and metallicity difference between the calibrator sample and the target sample. She suggested that the turnover magnitudes of the GCLF vary from galaxy to galaxy at the level of $\pm0.2$ mag because the intrinsic dispersion of the turnover magnitudes depends on the sampled globular cluster system in a galaxy. Therefore, the systematic error for galaxy-to-galaxy GCLF turnover magnitude scatter due to population and sampling effects  is estimated to be $\pm0.2$ mag in the case of the turnover magnitude measurements based on the entire globular clusters in a galaxy.

If we use only the metal-poor globular clusters, the corresponding uncertainty will be much smaller than the error based on the entire globular cluster samples because the metal-poor globular cluster calibrations based on three different samples of galaxies in \citet{dic06} agree at the level of 0.01 mag. 
It is difficult to estimate a precise value of this error at the moment.  In this study we adopt the error, in a conservative manner, to be $\pm0.1$ mag.

Combining this error due to the intrinsic scatter ($\pm0.1$) and the systematic error of the $M_V$(max) calibration ($\pm0.10$),
we estimate the total systematic error for the GCLF distance estimation to be $\pm0.14$ mag in the case of the metal-poor globular cluster samples. The corresponding error will be $\pm0.22$ mag in the case of  the entire globular cluster samples.
We summarize these errors in Table \ref{tab_gclfdistance}. 

\subsection{GCLF Distance Estimation for NGC 4589}
Applying the adopted $V$-band calibration for the metal-poor globular clusters ($M_V({\rm max})=-7.66\pm0.14$) to the measured turnover magnitude ($V({\rm max})=24.48\pm0.09$ and $V({\rm max})_0 =24.40\pm0.10$), 
we derive the distance to NGC 4589: 
$(m-M)_0= 32.06\pm0.10{\rm (ran)}\pm0.15{\rm (sys)} =32.06\pm0.18$ ($d=25.8\pm2.2$ Mpc). 
For the calculation of the errors, we included all uncertainties due to foreground extinction correction, aperture correction, and transformation uncertainties, as summarized in {\color{blue}\bf Table \ref{tab_gclfdistance}}.

If we use this result for the calibration of the $I$-band magnitudes
for NGC 4589 globular clusters, we obtain 
$M_I {\rm (max)} = (23.45\pm0.07) - 0.042 - (32.06\pm0.18) = -8.65\pm0.19$ mag,
 where 0.042 is the value for the $I$-band foreground extinction. 
This value is similar to the calibration
adopted in \citet{lee16},
$M_I {\rm (max)} = -8.56\pm0.09$ mag.

\section{Discussion}

\subsection{The Origin of SN 2005cz and Ca-rich SNe Ib}

To explain the spectral features and the light curves of SN 2005cz, \citet{kaw10} suggested that the origin of SN 2005cz is a core-collapse supernova whose progenitor is a massive star at the low-mass end ($8-12 M_\odot$) in a binary system. They noted a possible relation between SN 2005cz  and the young stellar population in the nucleus of NGC 4589. From the analysis of the spectrum of the NGC 4589 nucleus, \citet{zha08} suggested that the nucleus is dominated by old 
low-mass stars (age $>10^{10}$ yr and mass $M<1 M_\odot$), 
with 11\% of the nucleus composed of young stars (age $=10^7-10^8$ yr and mass $M<10 M_\odot$). \citet{kaw10} pointed out that the progenitor of SN 2005cz might have come from these young stars in the nucleus.

Later, \citet{suh11} presented, from $GALEX$ $NUV$ and $SDSS$ photometry,  that the observed $(NUV-r)$ color of NGC 4589 is close to the color of a galaxy 
that had little recent star formation 
(see their Fig. 2). Considering the possible presence of internal extinction due to dust clouds, they inferred that the intrinsic color of NGC 4589 may be much bluer than the observed color, and they favored the massive-star origin of SN 2005cz. They suggested that the progenitor of SN 2005cz may have a mass of $5-6 M_\odot$, which is even lower than the value of $8-12 M_\odot$ suggested by \citet{kaw10}. 
However, there is no information on the estimated value of the internal extinction, so 
it is not possible to tell whether the observed color is intrinsically red or significantly reddened by  internal dust.
  
In contrast, \citet{per11} used optical spectroscopy, H$\alpha$ emission, $GALEX$ UV emission, and $HST$ images to search for any signatures of young stellar populations around SN 2005cz. In particular, they tried to find young massive stars ($M>15 M_\odot$) from the photometry of point sources around the SN location using 
$HST$/WFPC2 and ACS images. 
However, \citet{per11} found no feature of any  young stellar population either close to or far ($>1.5$ kpc) from the position of SN 2005cz, supporting the old low-mass star origin scenario of SN 2005cz. 
\citet{per11} disfavored the massive-star origin of SN 2005cz, and showed that the spectral energy distribution of NGC 4589 covering $GALEX$ FUV to $2MASS$ $K_s$ is fitted very well by an old galaxy model 
with an age of 12.5 Gyr, stellar mass of $10^{11.17}$ $M_\odot$, and no specific star formation rate.

In this study, we found a small population of young star clusters in the central region at $R<0\farcm5$ ($<3.8$ kpc) that includes the location of SN 2005cz. 
Most of these star clusters are slightly resolved (i.e., larger than point sources) so that they cannot be individual stars in NGC 4589. 
Their colors range from $(B-V)=0$ to 0.5 ($(V-I)=0.2$ to 0.7), which
indicates that they are younger than about $10^9$ yr. 
The magnitudes of these clusters are $25<V\leq27$ ($-7.1 <M_V \leq-5.1$) mag. 
These magnitudes correspond to the masses of %
$10^{3.5} - 10^{4.5} M_\odot$ for the Padova SSP models with solar metallicity. The mass range does not change much with the choice of metallicity for $Z \gtrsim 0.1Z_\odot$. 
These young clusters are found only in the central region at $R<0\farcm5$ ($<3.8$ kpc), while old globular clusters are found in a much wider area with a much higher abundance.
These young star clusters might have formed in relation with a recent merger that resulted in a rotating dust disk \citep{moe89}.
They might have provided a massive-star progenitor for SN 2005cz.
Therefore, our finding of young star clusters in the central region of NGC 4589 supports the massive-star progenitor scenario for the origin of Ca-rich SNe Ib.

\subsection{Comparison with the Previous GCLF}

Using  shallow $HST$/WFPC2 $F555W$ and $F814W$ images, \citet{kun01} derived a
GCLF for NGC 4589 from the sample of globular clusters with $21<V<24.5$ mag. 
They could not determine the value of the GCLF width for NGC 4589 because of the shallow photometric limit.
They fitted it with 
a Gaussian function for a fixed width of $\sigma=1.3$. They obtained turnover magnitudes, $V({\rm max})_0 =25.22\pm0.39$ and $I({\rm max})_0 =24.21\pm0.41$.
These values are $\sim$0.8 mag
fainter than those in this study, $V{\rm (max)}_0 = 24.45\pm0.07$ mag 
and $I{\rm (max)}_0 = 23.43\pm0.07$ mag. 
The turnover magnitudes  in \citet{kun01} are about 0.9 mag fainter than 50\% completeness limits of their photometry, $V_{\rm lim} =24.3$ mag and $I_{\rm lim} =23.2$ mag. Their photometry did not reach the turnover magnitudes and  the errors of their estimated magnitudes are as large as 0.4 mag. 
Note that our photometry reaches much deeper than the turnover magnitudes.
Therefore the differences between \citet{kun01} and this study are considered to be mainly due to the shallow photometry in \citet{kun01}.  

\citet{kun01} also presented the total number of globular clusters, $N(total)=789\pm123$,
and the specific frequency, $S_N=5.1\pm3.7$. 
In this estimation, they adopted $(m-M)_0=31.95$ and $M_V=-21.2$ mag for NGC 4589.
However this value is a local $S_N$, constructed from a shallow photometry of asample based on only a small fraction of the entire galaxy, so it cannot be compared directly with our estimate. 

\begin{deluxetable*}{lcl}
\tabletypesize{\footnotesize}
\setlength{\tabcolsep}{0.05in}
\tablecaption{A List of Previous Distance Estimates for NGC 4589}
\tablewidth{0pt}
\tablehead{ \colhead{Method} & \colhead{Distance Modulus (Distance)}  & \colhead{Reference}}
\startdata
$D_n-\sigma$ & $33.91\pm0.25$ (60.6 Mpc) & \citet{fab89} \\
$D_n-\sigma$  & $32.93\pm0.40$ (38.6 Mpc) & \citet{wil97} \\
Faber-Jackson & $31.60\pm0.39$ (20.9 Mpc) & \citet{dev84} \\
Tully-Fisher & $32.55\pm0.40$ (32.3 Mpc) &  \citet{the07} \\
SBF & $31.71\pm0.22$ (22.6 Mpc) &  \citet{ton01}\\
SBF & $31.55\pm0.22$ (20.4 Mpc) & \citet{jen03}, updated \citet{ton01}-$I$-band\\
SBF & $31.77\pm0.24^a$ (20.4 Mpc) & \citet{bla10}, updated \citet{ton01}-$I$-band \\
SBF & $32.07\pm0.13^a$ (25.9 Mpc) & \citet{bla10}, updated \citet{jen03}-$F160W$-band \\
\hline
GCLF & $32.06\pm0.18$ (25.8 Mpc) & This Study \\
\hline
\enddata
\label{tab_distancelist}
\tablenotetext{a}{ 
The errors include the systematic $\pm0.1$ mag error in the optical and near-IR SBF measurements \citep{bla10,can18}. 
}
\end{deluxetable*}

\subsection{Comparison with Previous Distance Estimates}

In {\color{blue}\bf Table \ref{tab_distancelist}}, 
we list the previous estimates of the distance to NGC 4589 based on various methods.
The distances based on the $D_n - \sigma$ relation, the Faber-Jackson relation, and the Tully-Fisher relation show a large scatter ($(m-M)_0 = 31.60$ to 33.91) and have large errors from 0.25 to 0.40 mag \citep{fab89,wil97,dev84,the07}.
The distance in this study,  
$(m-M)_0 =32.06\pm0.18$ ($d = 25.8\pm2.2$ Mpc) 
is in the middle of these values.

In contrast, the distances based on the SBF have smaller errors than the others.
\citet{ton01} presented a distance of $(m-M)_0 =31.71\pm0.22$ based on the $I$-band SBF. This value was updated to a 0.16 mag smaller value later by \citet{jen03}, $(m-M)_0 =31.55\pm0.22$,  who used the \citet{uda99} Cepheid Period-Luminosity relation for the calibration of the method. 
However, metallicity correction for Cepheid distances needs to be taken into account for better distance estimation.
Considering this, \citet{bla10} published a correction formula to \citet{ton01} distances (see Appendix A of the paper). The appendix also contains some clarifications on the differences between the SBF distances by \citet{ton01} and by \citet{jen03}.
According to this correction, 
the revised distance modulus to NGC 4589 is derived to be $(m-M)_0 =31.77\pm0.22$ mag.

\citet{jen03} also presented near-IR SBF measurements of NGC 4589. 
 The distance modulus to NGC 4589 based on the near-IR SBF measurements 
 can be derived using 
the $\bar{m}_{F160W}$ value in their
Table 2 (column 4) 
and $\bar{M}_{F160W}$ value in their eq. (1). 
We derive a distance modulus,
$(m-M)_0 =27.21(\pm0.08)+4.76(\pm0.03)=31.97\pm0.09$ mag. 
Following the suggestion in \citet{bla10}, we apply the correction for the metallicity-dependent Cepheid zero-point of \citet{ton01}, 0.1 mag, to this value  and we obtain $(m-M)_0 = 32.07\pm0.09$.

All the errors in the above SBF distances are random errors.
The systematic error of each of the optical and the near-IR SBF distances is $\sim\pm0.1$ mag based on the Cepheid zero-point error \citep{jen03,can18}.
Considering both the random errors and the systematic errors,  the distance moduli based on optical and near-IR SBF measurements to NGC 4589 are
$(m-M)_0 {\rm (optical~ SBF)} =31.77\pm0.24$, and  $(m-M)_0 {\rm (NIR~ SBF)} = 32.07\pm0.13$, respectively.

Thus, the near-IR  SBF distance modulus is $0.3\pm0.3$ mag larger than the optical SBF value. 
However, the difference is similar to its error so that both values agree at the level of one sigma. 
The distance in this study, $(m-M)_0 =32.06\pm0.18$, 
is 0.3 mag larger than the revised optical SBF distance, 
$(m-M)_0 =31.77\pm0.24$,  but is in excellent agreement with the near-IR SBF distance, 
$(m-M)_0 = 32.07\pm0.13$ \citep{jen03,bla10}. 

\section{Summary and Conclusion}
Using deep $HST$/ACS images, we detected a significant population of star clusters in NGC 4589. This population is dominated by old globular clusters, but includes a small population of young star clusters in the central region. We present $BVI$ photometry of these clusters in the Vega magnitude system. 
Main results are summarized as follows. 

\begin{enumerate}

\item We found a small population of young star clusters with age $< 10^9$ yr  in the central region at $R<0\farcm5$ ($<3.8$ kpc) of NGC 4589. These young clusters might have provided a massive star progenitor for SN 2005cz, which supports  the massive star origin scenario of Ca-rich SNe Ib. 

\item The color distribution of the globular clusters is clearly bimodal. 
GMM analysis with a homoscedastic option identifies a blue peak at $(B-V)=0.78\pm0.01$ ($(V-I)=1.00\pm0.01$) 
and a red peak at $(B-V)=0.99\pm0.02$ ($(V-I)=1.20\pm0.01$). 
With a heteroscedastic option, we find the blue and red peaks at similar colors: $(B-V)=0.73\pm0.01$ ($(V-I)=0.99\pm0.02$) 
and $(B-V)=0.93\pm0.03$ ($(V-I)=1.18\pm0.02$).

\item The radial number density profile
of the globular clusters is flatter than the surface brightness profile of the stellar light. The former is fitted by a S{\'e}rsic law with $n_{\rm AGC}=1.41\pm0.34$, while the latter is fitted well by a S{\'e}rsic law with 
$n_{\rm galaxy}=3.11\pm0.21$, which is close to the de Vaucouleurs profile.

\item The GCLFs are fitted well by a Gaussian function. The $V$-band turnover magnitudes for all globular clusters,  the blue globular clusters, and the red globular clusters are found to be similar: 
$V_{\rm AGC}({\rm max})=24.53\pm0.06$ mag for all globular clusters, 
$V_{\rm BGC}({\rm max})=24.48\pm0.09$ mag for the blue globular clusters, and 
$V_{\rm RGC}({\rm max})=24.58\pm0.10$ mag for the red globular clusters.
The corresponding $V$-band widths are also found to be similar: 
$\sigma_{\rm AGC}=0.96\pm0.05$, 
$\sigma_{\rm BGC}=0.96\pm0.07$, 
and  $\sigma_{\rm RGC}=1.05\pm0.10$. 

\item We derived the total number of globular clusters to be $N(total)= 640\pm50$,
and the specific frequency to be $S_N=1.7\pm0.2$.

\item 
Considering the calibration errors of the turnover magnitudes ($\pm0.1$ mag) and the intrinsic variation depending on their host galaxies and environment ($\pm0.1$ mag), we estimate the total systematic error for the GCLF distance measurement to  be $\pm0.14$ mag in the case of the metal-poor globular cluster sample, and $\pm0.22$ mag in the case of the entire globular cluster sample. 

\item Adopting the calibration for the metal-poor globular clusters \citep{dic06,rej12}, we derive a distance to NGC 4589 from the turnover magnitude of the blue globular clusters:
$(m-M)_0 =32.06\pm0.18$ and $d=25.8\pm2.2$ Mpc.
The distance modulus in this study is in excellent agreement with the near-IR SBF distance, 
$(m-M)_0 = 32.07\pm0.13$, but 
0.3 mag larger than the revised optical SBF distance, 
$(m-M)_0 =31.77\pm0.24$ \citep{jen03,bla10}. 
\end{enumerate}

The authors like to thank the referee for careful and useful suggestions that helped to 
clarify some confusing issues such as the SBF distances and helped to improve the original manuscript.
The authors are also grateful to Marina Rejkuba for clarifying the uncertainties of the GCLF distances. 
We thank Brian Cho for improving the English in the manuscript.
This work was supported by the National Research Foundation of Korea (NRF) grant
funded by the Korean Government (NRF-2017R1A2B4004632).
J.K. was supported by the Global Ph.D. Fellowship Program (NRF-2016H1A2A1907015) of the National Research Foundation.

\clearpage


\clearpage

\end{document}